\documentclass[12pt]{iopart}
\usepackage{graphicx}
\usepackage{amsfonts}
\usepackage{url}
\usepackage{epstopdf}
\usepackage{color}
\usepackage{tikz-cd}
\usepackage{comment}
\usepackage[colorlinks=true,linkcolor=blue,citecolor=red]{hyperref}%
\usepackage{bm}
\usepackage{amssymb}
\usepackage{cite}

\expandafter\let\csname equation*\endcsname=\relax
\expandafter\let\csname endequation*\endcsname=\relax
\usepackage{amsmath}

\def\E{{\mathbb E}}
\def\P{{\mathbb P}}
\def\R{{\mathbb R}}

\def\A{{\mathcal A}}

\def\C{{\mathcal C}}

\def\x{\bm{x}}
\def\r{\bar{r}}

\def\erf{\mathrm{erf}}

\def\ctanh{\mathrm{ctanh}}

\begin{document}

\title{Survival in a nanoforest of absorbing pillars}

\author{Denis~S.~Grebenkov}
 \ead{denis.grebenkov@polytechnique.edu}

\address{
Laboratoire de Physique de la Mati\`{e}re Condens\'{e}e, \\ 
CNRS, Ecole Polytechnique, Institut Polytechnique de Paris, 91120 Palaiseau, France}

\author{Alexei T. Skvortsov}
\address{
Maritime Division, Defence Science and Technology Group, 506 Lorimer Street, Fishermans Bend, Victoria 3207, Australia}

\date{\today}

\begin{abstract}
We investigate the survival probability of a particle diffusing
between two parallel reflecting planes toward a periodic array of
absorbing pillars.  We approximate the periodic cell of this system by
a cylindrical tube containing a single pillar.  Using a mode matching
method, we obtain an exact solution of the modified Helmholtz equation
in this domain that determines the Laplace transform of the survival
probability and the associated distribution of first-passage times.
This solution reveals the respective roles of several geometric
parameters: the height and radius of the pillar, the inter-pillar
distance, and the distance between confining planes.  This model
allows us to explore different asymptotic regimes in the probability
density of the first-passage time.  In the practically relevant case
of a large distance between confining planes, we argue that the mean
first-passage time is much larger than the typical time and thus
uninformative.  We also illustrate the failure of the capacitance
approximation for the principal eigenvalue of the Laplace operator.
Some practical implications and future perspectives are discussed.
\end{abstract}





\submitto{\JPA}

\maketitle

\section{Introduction}

When a particle diffuses through a complex environment filled with
traps, its survival probability, which determines the first-passage
time (FPT) distribution, depends on the geometric configuration in a
very sophisticated way
\cite{Redner,Rice,benAvraham,Metzler,Lindenberg,Condamin07,Grebenkov07,Benichou10b,Benichou11,Hofling13,Bressloff13,Grebenkov13,Grebenkov20}.
Most former theoretical studies were focused on the {\it mean} FPT or,
equivalently, on the overall reaction rate on that traps (see
\cite{Weiss86,Benichou08,Holcman14,Benichou14,Holcman,Grebenkov16,Guerin16} and
references therein).  Despite the impressive progress in understanding
the mean FPT for various stochastic processes, its dominant role as a
{\it unique} timescale determining the whole distribution has been
questioned \cite{Mattos12,Godec16a,Godec16b,Grebenkov18c}.  In fact,
even though the mean FPT characterizes well the diffusive exploration
of a bounded confining domain, the absorption or reaction event may
occur on much shorter time scales.  For instance, in the
physiologically relevant example of calcium diffusion towards
calcium-sensing receptors inside a presynaptic bouton, the mean FPT is
usually around tens of millisecond, whereas the typical FPT is two or
even three orders of magnitude shorter \cite{Reva21}.  The limited
role of the mean FPT is particularly clear for unbounded domains, for
which the mean FPT is infinite due to a large contribution of rare
long trajectories.  The whole distribution of the FPT is therefore
required for a systematic comprehension of diffusion-controlled
reactions and related search processes.

For this purpose, many efforts were dedicated to characterize the
long-time behavior of the survival probability in disordered or random
environments \cite{benAvraham,Hughes,Levernier19} such as random packs
of absorbing immobile spheres \cite{Kayser83,Kayser84,Torquato91},
near a fractal boundary \cite{Levitz06}, or in dynamic heterogeneous
media \cite{Lanoiselee18}.  The random trajectories that survived up
to long times thoroughly explore the confining environment and thus
keep some averaged information on its geometric structure.  Their
contribution to the survival probability determines the right tail of
the probability density function (PDF) of the FPT.  In turn, the
short-time behavior of the survival probability is controlled by
so-called ``direct trajectories'' which are close to the shortest
geodesic path between the starting point and the closest trap
\cite{Godec16a,Basnayake18,Grebenkov22d}.  Such trajectories are
therefore sensitive only to the local geometric structure, yielding
rather universal short-time behavior in the left tail of the PDF.  Its
mathematical description goes back to the seminal works by Varadhan
\cite{Varadhan67a,Varadhan67b} and resembles the concepts of geometric
optics in physics \cite{Smith19,Meerson22}.

In contrast, the whole distribution of the FPT, that encompasses all
time- and geometric lengthscales, is known exactly only for rather
simple configurations such as an interval, a rectangle, a disk, a
sphere, or a pair of coaxial cylinders or concentric spheres
\cite{Redner,Carslaw,Crank,Thambynayagam}.  In these settings, the
symmetry of the confining domain allows for a separation of variables
and leads to explicit representations of the survival probability and
the PDF of the FPT.  When the absorbing region is only a part of the
otherwise reflecting boundary, such basic methods do not work anymore,
and one has to employ more sophisticated tools.  For instance,
Isaacson and Newby proposed a uniform in time asymptotic expansion for
the PDF of the FPT to a small target \cite{Isaacson13}.  Another
approach was used in Ref. \cite{Rupprecht15} to compute the survival
probability inside two-dimensional rotationally invariant domains
(like a disk or a sector) in the presence of an absorbing arc on the
boundary.  Both an exact solution relying on a matrix inversion, and
an approximate explicit solution were proposed.  In the case of
domains formed by coaxial cylinders or concentric spheres, the
survival probability in the presence of an absorbing region was
obtained with the aid of the self-consistent approximation
\cite{Grebenkov18d,Grebenkov19c,Grebenkov21c}.  A general method for
getting the survival probability in a medium with multiple spherical
traps was described in \cite{Grebenkov20d}.

\begin{figure}
\begin{center}
\includegraphics[width=50mm]{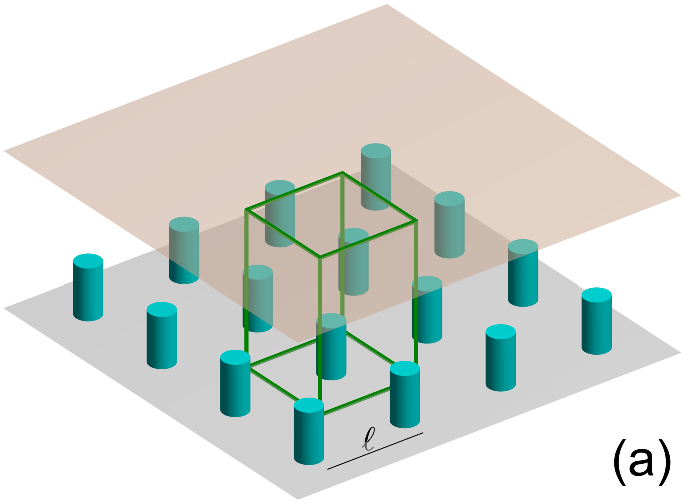} 
\includegraphics[width=25mm]{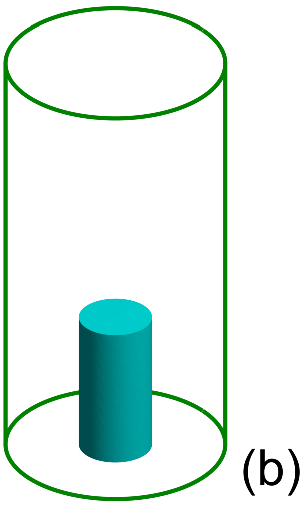} 
\includegraphics[width=25mm]{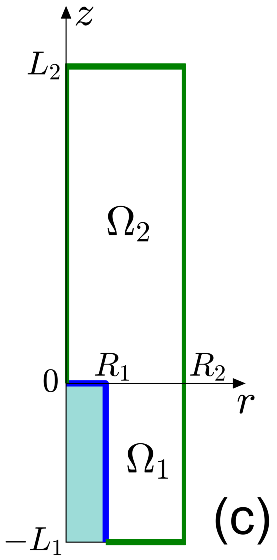} 
\includegraphics[width=50mm]{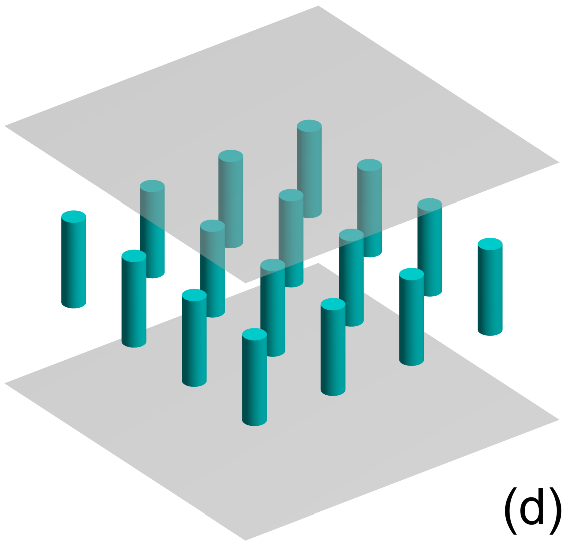} 
\end{center}
\caption{
{\bf (a)} A square-lattice array of cylindrical pillars (in light
blue) on a reflecting support (in gray), capped by an upper reflecting
plane (in pink).  Periodicity of this domain allows one to focus
on diffusion in a periodic cell around one pillar -- a green
rectangular parallelepiped. {\bf (b)} A single pillar surrounded by an
{\it effective} coaxial reflecting cylindrical tube and capped by two
parallel reflecting planes.  {\bf (c)} Planar ($xz$) projection of the
three-dimensional domain from panel {\bf (b)}.  Blue segments show the
absorbing pillar and green segments represent reflecting parts 
(the green vertical segment at $r = 0$ is also reflecting to respect
the regularity and the axial symmetry of the solution, see
\ref{sec:derivation}).  Shadowed (light blue) region is the solid
(inaccessible) interior of the cylindrical pillar.  Here $R_2$ is the
radius of the outer reflecting cylinder, $R_1$ is the radius of the
absorbing pillar, $L_1$ is its height, and $L_2$ is the distance
between the source and the top of the pillar (i.e., $L_1+L_2$ is the
height of the whole system).  Note that $R_2$ is related to the
inter-pillar distance $\ell$, e.g., $R_2 = \ell/\sqrt{\pi}$ for the
square lattice.  {\bf (d)} An equivalent problem of twice longer
pillars between two reflecting planes separated by distance
$2(L_1+L_2)$.}
\label{fig:scheme}
\end{figure}

In a recent paper \cite{Grebenkov23}, we studied steady-state
diffusion from a remote source towards a periodic array of absorbing
identical cylindrical pillars protruding from a flat base
(Fig. \ref{fig:scheme}(a)).  Using a mode matching method
\cite{Grebenkov18,Delitsyn18,Delitsyn22}, we solved the underlying
Laplace equation and found the exact form of the diffusive flux onto
each pillar, $J = c_0 D \A/(L_2 + z_0)$, where $c_0$ is the imposed
concentration of particles at the source, $D$ is the diffusion
coefficient, $\A$ is the cross-sectional area of a periodic cell,
$L_2$ is the distance between the source and the top of the pillars,
and $z_0$ is the offset parameter that aggregates the geometric
complexity and reactivity of the spiky coating.  Using the exact
though sophisticated expression for $z_0$, we analyzed the behavior of
the steady-state diffusive flux in different asymptotic regimes.

In the present work, we extend the above analysis to the modified
Helmholtz equation, $(p - D\Delta) u = 0$, which describes diffusion
in a reactive medium with the bulk reaction rate $p$; in addition,
this equation results from the Laplace transform of the diffusion
equation $\partial_t c = D\Delta c$ and thus gives access to
time-dependent diffusion.  In particular, we focus on the survival
probability of a particle diffusing towards a nanoforest of absorbing
pillars.  We obtain the exact solution for the Laplace transform of
this quantity that yields the moments and the PDF of the FPT to
absorbing pillars.

The paper is organized as follows.  In Sec. \ref{sec:general}, we
formulate the problem and describe the main steps of its solution.
Section \ref{sec:discussion} presents several properties of the FPT
distribution.  In particular, we discuss the short-time and long-time
asymptotic behaviors of the PDF of the FPT, the limited significance
of the mean FPT, the failure of the capacitance approximation for the
decay time, and the respective roles of different geometric parameters
of the nanoforest.  Conclusions and open problems are summarized in
Sec. \ref{sec:conclusion}.  Details of the derivation are re-delegated
to Appendices.

\section{Exact solution}
\label{sec:general}

We consider ordinary diffusion of a point-like particle between two
reflecting planes at $z = -L_1$ and $z = L_2$ in three dimensions.
The bottom plane is covered by a square-lattice array of absorbing
identical cylindrical pillars of radius $R_1$ and height $L_1$, with
the inter-pillar distance $\ell$ between the centers of the closest
pillars (Fig. \ref{fig:scheme}(a)).  The periodicity of this array
allows one to focus on diffusion in a periodic cell containing a
single pillar, i.e., inside a rectangular parallelepiped
$(-\ell/2,\ell/2)^2 \times (-L_1,L_2)$ with periodic boundary
conditions along $x$ and $y$ directions.  Following the rationale by
Keller and Stein \cite{Keller_1967}, we replace this original periodic
cell by a cylindrical tube with reflecting boundary condition.  The
radius $R_2$ of the tube is chosen to preserve the volume of the
periodic cell, i.e., by setting $\pi R_2^2 = \ell^2$.  In this way, we
will {\it approximate} the solution of the original problem by the
exact solution of the reduced problem.  The accuracy of this
approximation can be accessed by a numerical solution of the original
problem.  Its systematic study will be presented elsewhere (see the
related discussion in \cite{Cai92} for the Laplace equation in a
different geometric setting).

From now on, we focus on diffusion inside a bounded domain $\Omega$,
surrounded by a cylindrical tube of radius $R_2$, towards a co-axial
cylindrical absorbing pillar of radius $R_1$ and height $L_1$, both
confined between parallel reflecting planes at $z = -L_1$ and $z =
L_2$ (Fig. \ref{fig:scheme}(b)).  Starting from a point $\x$ inside
this confining domain, the particle moves with the diffusion
coefficient $D$ until the first arrival onto the surface of the
pillar.  The first-passage time to that surface, $\tau$, is a random
variable, which is fully characterized by the survival probability,
$S(t|\x) = \P\{ \tau > t\}$.  The latter satisfies the (backward)
diffusion equation, $\partial_t S = D\Delta S$, which is supplemented
by the initial condition $S(0|\x) = 1$ and mixed boundary conditions:
$S(t|\x) = 0$ on the absorbing pillar, and $\partial_z S(t|\x) = 0$ on
the reflecting planes.  The negative time derivative, $H(t|\x) =
-\partial_t S(t|\x)$, is the probability density function of the FPT
$\tau$.  In turn, the Laplace transform of the survival probability,
\begin{equation}
\tilde{S}(p|\x) = \int\limits_0^\infty dt \, e^{-pt} \, S(t|\x),
\end{equation}
satisfies the modified Helmholtz equation, subject to the same
boundary conditions.  More explicitly, we search for the
Laplace-transformed survival probability that satisfies the following
boundary value problem in cylindrical coordinates $\x = (r,z,\phi)$:
\begin{subequations}  \label{eq:problem_S}
\begin{align}  \label{eq:Helm_S}
& (p - D\Delta) \tilde{S} = 1  \quad \textrm{in}~\Omega, \\  
\label{eq:cond_disk_S}
& \tilde{S} = 0 \quad (r < R_1,~ z = 0),\\  
\label{eq:cond_inner_S}
& \tilde{S} = 0 \quad (r = R_1,~ -L_1 < z < 0),\\  
\label{eq:cond_top_S}
& \partial_z \tilde{S} = 0 \quad (0 < r < R_2,~ z = L_2), \\  
\label{eq:cond_bottom_S}
& \partial_z \tilde{S} = 0 \quad (R_1 < r < R_2,~ z = -L_1), \\
\label{eq:cond_outer_S}
& \partial_r \tilde{S} = 0 \quad (r = R_2,~ -L_1 < z < L_2),    
\end{align}
\end{subequations}
where $\Delta = \partial_r^2 + (1/r) \partial_r + \partial_z^2$ is the
Laplace operator in cylindrical coordinates (without the angular
part).  Here, Eqs. (\ref{eq:cond_disk_S}, \ref{eq:cond_inner_S})
incorporate absorption on the pillar, while Eqs. (\ref{eq:cond_top_S},
\ref{eq:cond_bottom_S}, \ref{eq:cond_outer_S}) describe reflections of
the particle on the top and bottom boundaries and on the outer
cylindrical surface.  The rotation invariance of this problem implies
that $\tilde{S}(p|r,z)$ does not depend on the angle $\phi$, which
therefore will be omitted in what follows.  Note also that the
reflection with respect to the plane at $z = -L_1$ transforms this
geometric setting into an equivalent one, with a twice longer pillar
located in the middle of the cylindrical tube of height $2(L_1+L_2)$.
In other words, we also approximate the Laplace-transformed survival
probability in the presence of twice longer absorbing pillars located
in the middle between two parallel reflecting planes
(Fig. \ref{fig:scheme}(d)).

Setting $\tilde{S}(p|r,z) = (1 - \tilde{H}(p|r,z))/p$, one can
transform the above inhomogeneous modified Helmholtz equation into the
homogeneous one:
\begin{subequations}  \label{eq:problem_u}
\begin{align}  \label{eq:Helm}
& (p - D\Delta) \tilde{H} = 0  \quad \textrm{in}~\Omega, \\  
\label{eq:cond_disk}
& \tilde{H} = 1 \quad (r < R_1,~ z = 0),\\  
\label{eq:cond_inner}
& \tilde{H} = 1 \quad (r = R_1,~ -L_1 < z < 0),\\  
\label{eq:cond_top}
& \partial_z \tilde{H} = 0 \quad (0 < r < R_2,~ z = L_2), \\  
\label{eq:cond_bottom}
& \partial_z \tilde{H} = 0 \quad (R_1 < r < R_2,~ z = -L_1) ,\\
\label{eq:cond_outer}
& \partial_r \tilde{H} = 0 \quad (r = R_2,~ -L_1 < z < L_2).
\end{align}
\end{subequations}
In this way, we focus directly on the Laplace transform
$\tilde{H}(p|r,z)$ of the PDF $H(t|r,z)$ of the FPT.  Since
$\tilde{H}(p|r,z) = \E\{ e^{-p\tau}\}$, the derivatives of this
function with respect to $p$ determine the integer-order moments of
the FPT:
\begin{equation}
\E\{ \tau^k \} = (-1)^k \lim\limits_{p\to 0} \frac{\partial^k \tilde{H}(p|r,z)}{\partial p^k}  \,.
\end{equation}
Moreover, the function $\tilde{H}(p|r,z)$ admits another
interpretation as a steady-state concentration of particles, emitted
from the pillar into a reactive medium with the bulk reactivity $p$.
Yet another probabilistic interpretation is that $\tilde{H}(p|r,z)$ is
the probability for a particle started from $\x = (r,z,\phi)$ to
arrive onto the pillar before being killed in the bulk.  In other
words, it describes the survival of a mortal random walker 
\cite{Yuste13,Meerson15,Grebenkov17d,Meerson19}.

In \ref{sec:exact}, we derive the exact solution of the problem
(\ref{eq:problem_u}) by using a mode matching method
\cite{Grebenkov18,Delitsyn18,Delitsyn22,Grebenkov23}.  In a nutshell,
one represents a general solution of Eq. (\ref{eq:Helm}) in subdomains
with $z < 0$ and $z > 0$ as two series (\ref{eq:u1}, \ref{eq:u2})
involving appropriate Bessel functions.  The continuity and
differentiability of the solution at the junction $z = 0$ imply an
infinite system (\ref{eq:system}) of linear algebraic equations on the
unknown coefficients of these series.  The elements of the
infinite-dimensional matrix $W$ that defines this system, are known
{\it explicitly} through Eq. (\ref{eq:Wdef}).  Truncating this system
to a finite size $N$, one can solve it numerically by inverting a
finite-size matrix.  Despite the need for a numerical step, the
obtained solution provides an analytic dependence of
$\tilde{H}(p|r,z)$ on the coordinates $r$ and $z$ of the starting
point.  Moreover, the truncation error rapidly decreases with $N$,
allowing one to use moderate truncation orders (say, few tens) and
thus very rapid computations for a broad range of parameters.
Finally, the structure of the solution reveals the respective roles of
different parameters and opens a way to asymptotic analysis.  In the
following, we mainly focus on the PDF $H(t|r,z)$ that can be obtained
numerically by representing the inverse Laplace transform of
$\tilde{H}(p|r,z)$ as the Bromwich integral and approximating it with
the help of the Talbot algorithm \cite{Talbot79}.  We fixed the
truncation size $N = 10$ and checked that this choice was sufficient
to get accurate results.

As diffusion occurs in a bounded domain, the survival probability and
the PDF of the FPT admit general spectral expansions
\begin{equation}  \label{eq:St_expansion}
S(t|r,z) = \sum\limits_{n=0}^\infty e^{-Dt\lambda_n} \, u_n(r,z) \, \int\limits_\Omega d\x \, u_n(\x) 
\end{equation}
and
\begin{equation}  \label{eq:Ht_expansion}
H(t|r,z) = \sum\limits_{n=0}^\infty D\lambda_n \, e^{-Dt\lambda_n} \, u_n(r,z) \, \int\limits_\Omega d\x \, u_n(\x) ,
\end{equation}
where $\lambda_n$ and $u_n(\x)$ are the eigenvalues and
$L_2(\Omega)$-normalized eigenfunctions of the (negative) Laplace
operator $-\Delta$.  The eigenvalues, which are positive and
enumerated in an ascending order, are determined by the poles
$\{p_n\}$ of $\tilde{S}(p|r,z)$ as $\lambda_n = - p_n/D$.  In turn,
the poles are obtained as the values of $p$ in the complex plane
${\mathbb C}$, at which the matrix $I+W$ is not invertible, i.e., when
$\det(I+W) = 0$ (with $I$ being the identity matrix).  As the
eigenvalues are positive, one can search for the poles $p_n$ on the
negative axis (see details in \ref{sec:poles}).  In turn, the
eigenfunctions and the coefficients (given by the integral) are
determined from the residues of $\tilde{S}(p|r,z)$ at the poles.
Despite the simple intuitively appealing form of these spectral
expansions, their numerical computation is tedious so that we
performed a numerical inversion of the Laplace transforms
$\tilde{S}(p|r,z)$ and $\tilde{H}(p|r,z)$, as described above.

\section{Discussion}
\label{sec:discussion}

In this section, we discuss the properties of the survival probability
$S(t|r,z)$ and the PDF $H(t|r,z)$ of the FPT.  In particular, we aim
at understanding the respective roles of different geometric
parameters of the system, namely, the pillar's radius $R_1$ and height
$L_1$, the distance $L_2$ to the top reflecting plane, and the radius
$R_2$ of the outer reflecting surface, which is related to the
inter-pillar distance $\ell$.  Throughout this discussion, we fix the
radius $R_2$ and rescale all other lengths by $R_2$.  While the
obtained exact solution is valid for any set of these parameters, we
will mainly focus on configurations, in which $L_2/R_2$ is large and
$\rho = R_1/R_2$ is small.  In all numerical examples, we set $R_2 =
1$ and $D = 1$ to fix units of length and time.

We generally discuss the whole distribution of the FPT and its
asymptotic behaviors.  As said earlier, the short-time asymptotic
behavior is determined by ``direct trajectories'' that go straight
from the starting point to the closest point on the pillar
\cite{Godec16a,Godec16b}.  As a consequence, the left tail of the PDF
is very sensitive to the starting point and to the closest part of the
pillar.  In turn, the geometric configuration of the system does not
almost affect this behavior.  As earlier discussed for other settings
\cite{Grebenkov18c,Grebenkov18d,Grebenkov19c,Grebenkov21c}, one
generally gets the L\'evy-Smirnov type behavior,
\begin{equation}
H(t|r,z) \sim  \frac{\delta}{\sqrt{4\pi D t^3}} e^{-\delta^2/(4Dt)} \qquad (t\to 0),
\end{equation}
where $\delta$ is the distance between the starting point and the
absorbing pillar.  As this short-time behavior is rather universal, we
do not dwell on its analysis.  In contrast, we focus on the
intermediate- and long-time behaviors when the particle has enough
time to explore the bulk around the pillar and is thus sensitive to
the geometric configuration of the system.

\subsection{Long-time behavior}
\label{sec:decay-time}

The spectral expansion (\ref{eq:Ht_expansion}) implies an exponential
decay of the PDF at long times:
\begin{equation}  \label{eq:Ht_exp}
H(t|r,z) \approx \frac{e^{-t/T}}{T} \, u_0(r,z) \, \int\limits_\Omega d\x \, u_0(\x) ,
\end{equation}
where the decay time $T = 1/(D\lambda_0)$ is determined by the
principal (smallest) eigenvalue $\lambda_0$, which depends on the
geometric parameters of the domain $\Omega$ is a sophisticated way.

To get some insights onto the decay time, let us first establish a
simple upper bound.  If the starting point $\x$ is located in the
upper part of the domain (with $z > 0$), the survival probability
obeys the following inequality
\begin{equation}  \label{eq:S_ineq1}
S(t|r,z) \geq S_1(t|z)  \qquad (t \geq 0, ~ z > 0),
\end{equation}
where $S_1(t|z)$ is the survival probability in a capped cylinder of
radius $R_2$ with an absorbing disk at $z = 0$ and a reflecting disk
at $z = L_2$.  Due to the axial symmetry, this is actually the
survival probability on the interval $(0,L_2)$ with the absorbing
endpoint $0$ and the reflecting endpoint $L_2$.  This inequality
follows from the continuity of Brownian motion: any trajectory that
hits the absorbing pillar at time $t$ should cross the level $z = 0$
and thus hit the disk at an earlier time $t'$, i.e., it is more
probable to avoid the contact with the pillar than the contact with
the absorbing disk at $z = 0$.  The survival probability $S_1(t|z)$ is
known explicitly (see, e.g., \cite{Redner}) and is reproduced in
Eq. (\ref{eq:St_interval}) for completeness.  In particular, it decays
exponentially at long times, with the decay rate $D\pi^2/(4L_2^2)$.
To ensure the inequality (\ref{eq:S_ineq1}), the decay rate of
$S(t|r,z)$ should be slower than (or equal to) the decay rate of
$S_1(t|z)$, i.e., $\lambda_0 \leq \pi^2/(4L_2^2)$.  Similarly, if the
particle starts from a point with $r > R_1$, the survival probability
obeys another inequality
\begin{equation}  \label{eq:S_ineq2}
S(t|r,z) \geq S_2(t|r)  \qquad (t \geq 0,~ r > R_1),
\end{equation}
where $S_2(t|r)$ is the survival probability inside the annulus
between an absorbing circle of radius $R_1$ and a reflecting circle of
radius $R_2$.  Once again, before hitting the pillar, any trajectory
started from a point with $r > R_1$ must cross the cylindrical surface
at $r = R_1$, whatever the vertical coordinate is.  The survival
probability $S_2(t|r)$ in the annulus also admits an explicit solution
(see, e.g., \cite{Thambynayagam}) and is reproduced in
Eq. (\ref{eq:St_annulus}).  Its long-time behavior is determined by
the decay rate $D\alpha_{0,1}^2/R_2^2$ so that
$\lambda_0 \leq \alpha_{0,1}^2/R_2^2$, where $\alpha_{0,1}$ is the
smallest positive solution of Eq. (\ref{eq:alpha1}).  Combining two
inequalities, we get the following lower bound for the decay time
\begin{equation}  \label{eq:T_lower}
T \geq \max \biggl\{ \frac{R_2^2 }{\alpha_{0,1}^2 D},\, \frac{4L_2^2}{\pi^2 D} \biggr\} .
\end{equation}
Depending on the geometric parameters, either of two bounds can be
dominant.  If the pillar is very thin, $\alpha_{0,1}^2$ is small, so
that $R_2^2/(\alpha_{0,1}^2 D)$ can be the maximum, if $L_2/R_2$ is
not too large (see further discussion in Sec. \ref{sec:R1}).  In
contrast, if $L_2/R_2$ is large enough, $4L_2^2/(\pi^2 D)$ is the
maximum.  Since $\alpha_{0,1}^2$ decreases logarithmically slowly as
$R_1 \to 0$ according to Eq. (\ref{eq:alpha0_rho0}), the latter case
is more relevant for applications.  Note also that the upper bound
does not depend on the pillar's height $L_1$; one can therefore expect
that the impact of this geometric parameter onto the decay time is
moderate, at least in the settings with large $L_2/R_2$.  We return to
this point in Sec. \ref{sec:L1}.

\subsubsection{Mean FPT}

When the target is small (as compared to the confining domain), the
decay time $T$ is usually close to the mean FPT.  In the regime
$L_2/R_2 \gg 1$, there is a simple approximation for the mean FPT.  In
fact, a spiky bottom surface can be approximated by an effective
absorbing flat boundary located at $z = - z_0$, where the offset
parameter $z_0$ was thoroughly investigated in \cite{Grebenkov23}.  In
this way, the original problem is reduced to one-dimensional diffusion
on the interval $(-z_0,L_2)$ with the absorbing endpoint $-z_0$ and
the reflecting endpoint $L_2$, for which the mean FPT is
\begin{equation}
T(z) = \frac{(z+z_0)(2L_2 + z_0 - z)}{2D}  \,.
\end{equation}
Moreover, if the starting point is uniformly distributed, the volume
average of $T(z)$ yields $\overline{T} = (L_2+z_0)^2/(3D)$, where
$z_0$ incorporates the dependence on the geometric parameters of the
system.  As $z_0$ is usually much smaller than $L_2$, one has
$\overline{T} \approx \frac13 L_2^2/D$, which is close to the lower
bound $\frac{4}{\pi^2} L_2^2/D$ on the decay time $T$.

\subsubsection{Capacitance approximation}

When the pillar is small as compared to the cylindrical tube, i.e.,
$L_1/R_2 \ll 1$ and $\rho = R_1/R_2 \ll 1$, the reflecting boundary
can be treated as being at infinity, and one often approximates the
principal eigenvalue as
\cite{Mazya85,Ward93,Kolokolnikov05,Cheviakov11,Chaigneau22}
\begin{equation}  \label{eq:lambda0}
\lambda_0 \approx \frac{\C}{|\Omega'|} \,,
\end{equation}
where $|\Omega'| = 2|\Omega| = 2\pi (L_1(R_2^2-R_1^2) + L_2 R_2^2)$ is
the volume of the twice bigger domain $\Omega'$, which is obtained by
reflection with respect to the plane at $z = -L_1$, and $\C$ is the
capacitance of the twice longer pillar \cite{Sandua_2013}
\begin{equation}  \label{eq:Cpillar}
\C = 4\pi R_1 \frac{1 + (L_1/R_1)^2}{\frac{\pi}{2} + \frac{L_1}{R_1} \ln \frac{L_1}{R_1}}  
\end{equation}
(note that we use the convention, in which the capacitance of a sphere
of radius $r$ is $4\pi r$).  As a consequence, the capacitance
approximation (\ref{eq:lambda0}) implies the following expression for
the decay time
\begin{equation}
T_{\rm cap} = \frac{L_2 R_2}{2D} \, \frac{1 + (1-\rho^2) L_1/L_2}
{\rho \bigl(\frac{1 + (L_1/R_1)^2}{\pi/2 + (L_1/R_1) \ln (L_1/R_1)}\bigr) }  \,.
\end{equation}
In the same vein, the capacitance was employed to describe the mean
FPT, the overall reaction rate, and the long-time behavior of the
survival probability (see
\cite{Berg77,Berezhkovskii07,Lindsay_2017,Grebenkov22c} and references
therein).

However, one can see that this approximation is incompatible with the
lower bound (\ref{eq:T_lower}) in the regime $L_2/R_2 \gg 1$.  In
fact, the decay time $T_{\rm cap}$ grows linearly with $L_2$ when
other parameters are fixed, whereas the lower bound grows
quadratically with $L_2$.  This is a striking example of the failure
of the capacitance approximation (\ref{eq:lambda0}) for {\it
anisotropic} confining domains.  In other words, when speaking about
the small target limit, one has to take the double limit $L_2\to
\infty$ and $R_2\to \infty$ simultaneously to keep the confining
domain more or less isotropic.

\begin{figure}
\begin{center}
\includegraphics[width=100mm]{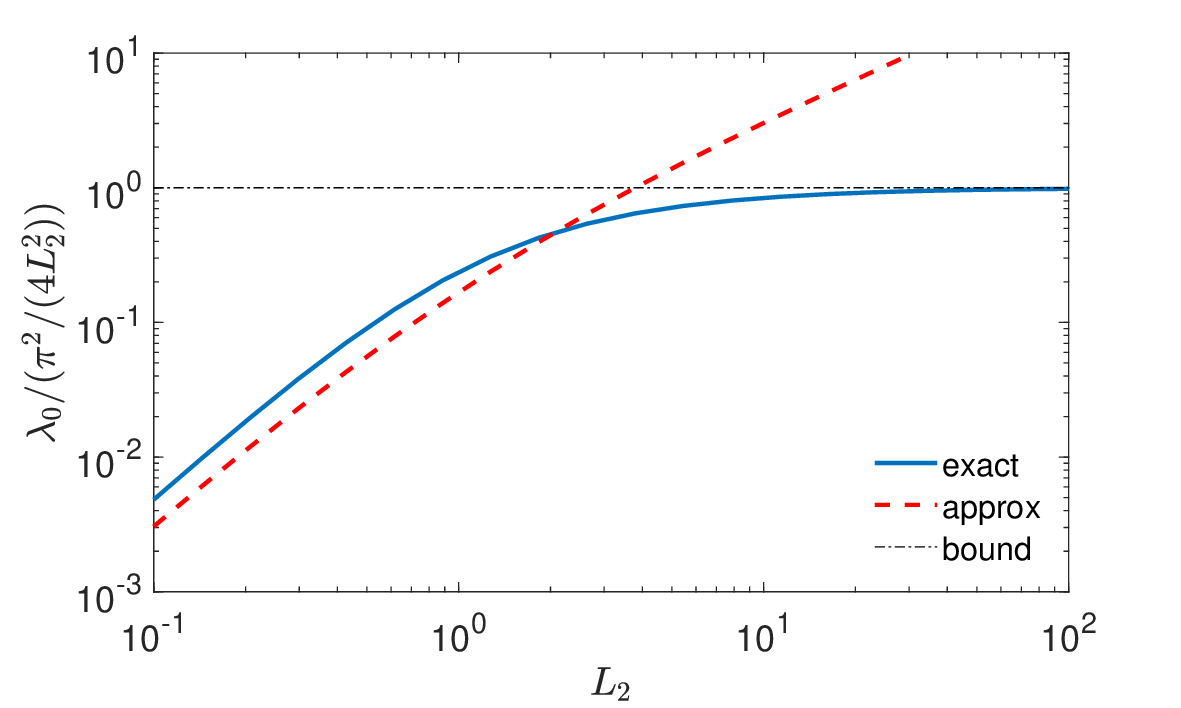} 
\end{center}
\caption{
Principal eigenvalue $\lambda_0$ of the Laplace operator, rescaled by
its upper bound $\pi^2/(4L_2^2)$, as a function of $L_2$, for the
domain with $R_1 = 0.1$, $L_1 = 1$ and $R_2 = 1$.  The eigenvalue
$\lambda_0$ (shown by solid line) was obtained as $-p_0/D$, where
$p_0$ is the pole of $\tilde{S}(p|r,z)$ with the smallest absolute
value, which was found numerically as the first zero of $\det(I+W) =
0$ (see \ref{sec:poles}).  Dashed line presents the capacitance
approximation (\ref{eq:lambda0}), while dash-dotted horizontal line
indicates the upper bound (here, it is located at $1$ due to
rescaling).}
\label{fig:lambda0}
\end{figure}

Figure \ref{fig:lambda0} illustrates the behavior of the principal
eigenvalue $\lambda_0$ as a function of $L_2$.  It is rescaled by
$\pi^2/(4L_2^2)$ to highlight the role of this upper bound.  One sees
that $\lambda_0$ rapidly approaches its upper bound as $L_2$
increases.  In turn, the capacitance approximation (\ref{eq:lambda0})
captures qualitatively the behavior of $\lambda_0$ when $L_2 \lesssim
2$ but then exceeds the upper bound and thus fails.  Note that the
shift between two curves at small $L_2$ is caused by the fact that the
target is not small enough as compared to the confining domain (here,
$R_1/R_2 = 0.1$ and $L_1/R_2 = 1$).  For smaller $R_1/R_2$ and/or
$L_1/R_2$ (not shown), the agreement in the region of small $L_2$ is
better, but the capacitance approximation still fails at large $L_2$.

\subsubsection{Role of the decay time}

We conclude that if $L_2/R_2$ is large, the decay time $T$ is close to
its lower bound $\frac{4}{\pi^2} L_2^2/D$.  Most importantly, it does
not almost depend on the geometric parameters of the system (except
$L_2$), i.e., this time scale is {\it uninformative} for the
considered first-passage process.  Similarly, the mean first-passage
time, which is usually close to the decay time, does not bear
substantial information on the search process in this case.  Moreover,
in the limit $L_2\to\infty$, the decay time and the mean FPT diverge
and therefore become useless.  For this reason, we do not discuss the
mean FPT in the remaining text and focus on the whole distribution.

\subsection{Role of distance $L_2$}
\label{sec:L2}

In many applications, the distance $L_2$ is much larger than the other
length scales.  An interesting question is how the long-time behavior
changes as $L_2$ goes to infinity.  In this limit, the principal
eigenvalue $\lambda_0$ vanishes so that the exponential decay
(\ref{eq:Ht_exp}) should transform into a slower decrease at $L_2 =
\infty$.  In the particular case $R_1 = R_2$, the original
three-dimensional problem reduces to one-dimensional diffusion on the
positive semi-axis $\R_+$, with the L\'evy-Smirnov PDF 
\begin{equation}  \label{eq:Ht_1D}
H_{\rm 1D}(t|z) = \frac{z e^{-z^2/(4Dt)}}{\sqrt{4\pi D t^3}} \,,
\end{equation}
behaving as $t^{-3/2}$ as $t\to \infty$ \cite{Redner}.  The origin of
this slow power-law decay is the existence of very long random
trajectories that can go arbitrarily far away from the absorbing point
at $z = 0$.  Even though such long trajectories are unlikely, their
contribution makes the mean FPT infinite.  The same probabilistic
argument holds in the case $R_1 < R_2$ so that the PDF $H(t|r,z)$
behaves as $t^{-3/2}$ in general.  In \ref{sec:long-time}, we deduce
this general behavior from the exact solution.  To grasp the origin of
this slow power-law decay without technical analysis, one can again
apply the inequality (\ref{eq:S_ineq1}), in which $S_1(t|z)$ is now
the survival probability on the positive semi-axis, which is known
exactly:
\begin{equation}  \label{eq:St_1D}
S_{\rm 1D}(t|z) = \erf\biggl(\frac{z}{\sqrt{4Dt}}\biggr),
\end{equation}
where $\erf(z)$ is the error function.  At long times, one has $S_{\rm
1D}(t|z) \approx z/\sqrt{\pi Dt}$ so that the survival probability
$S(t|r,z)$ cannot decrease faster than $t^{-1/2}$.  This simple
argument excludes, e.g., an exponential decay of $S(t|r,z)$ in the
limit $L_2 = \infty$.

\begin{figure}
\begin{center}
\includegraphics[width=100mm]{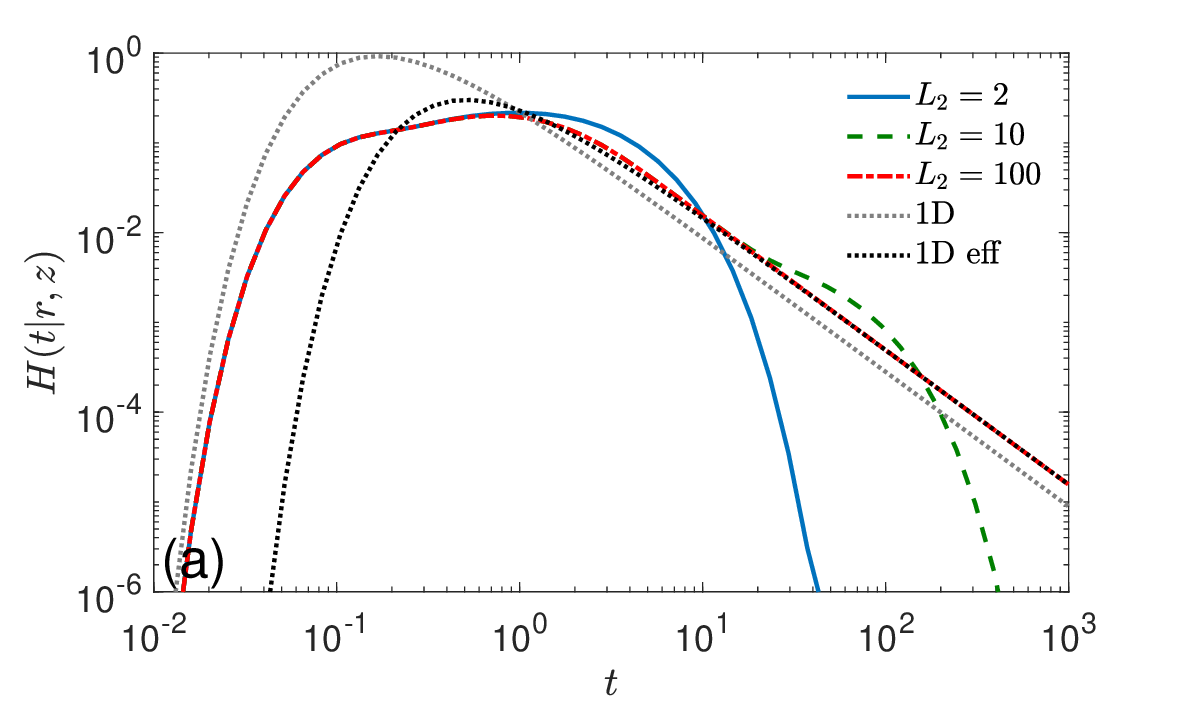} 
\includegraphics[width=100mm]{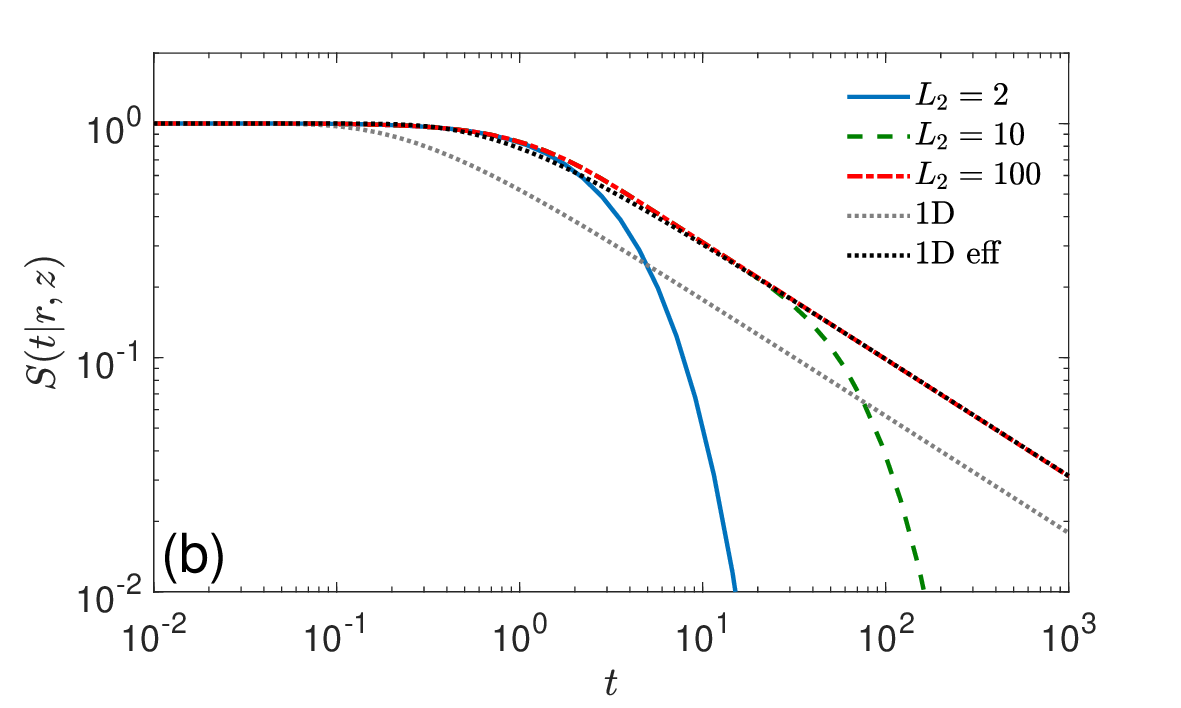} 
\end{center}
\caption{
{\bf (a)} Probability density $H(t|r,z)$ of the FPT to the absorbing
pillar, with $R_1 = 0.1$, $L_1 = 10$, $r = 0$, $z = 1$, and three
values of $L_2$ (see the legend).  Gray and black dotted lines present
respectively the probability densities $H_{\rm 1D}(t|r,z)$ and $H_{\rm
1D}(t|r,z+z_0)$ from Eq. (\ref{eq:Ht_1D}) for the half-line,
illustrating the emergence of a power-law shoulder before an
exponential cut-off.  The offset parameter $z_0 \approx 0.75$ was
calculated by the exact formula given in \cite{Grebenkov23}.  {\bf
(b)} Survival probability $S(t|r,z)$ and its approximations $S_{\rm
1D}(t|z)$ and $S_{\rm 1D}(t|z+z_0)$ given by Eq. (\ref{eq:St_1D}) for
the same setting.}
\label{fig:Ht_L2}
\end{figure}

Figure \ref{fig:Ht_L2}(a) illustrates the effect of an increasing
distance $L_2$ onto the probability density $H(t|r,z)$.  The starting
point is located above the top of the pillar, at a height $z = 1$.  At
short times, only ``direct'' trajectories to the pillar contribute to
the left tail of the PDF so that the distance $L_2$ to the top
boundary does not matter, and all three curves coincide.  In contrast,
the long-time limit corresponds to a diffusive exploration of the
bounded domain so that an increase of $L_2$ strongly affects the right
tail, shifting it to longer times.  Even though there is an
exponential cut-off for any finite $L_2$, one can clearly see the
emergence of an intermediate regime with a power-law decay $t^{-3/2}$,
starting from $t \gtrsim 1$, in agreement with the above analysis.
This behavior can be recognized by a straight line in the log-log
plot; for comparison, Eq. (\ref{eq:Ht_1D}) is also shown.  One sees
that the exponential cut-off is progressively shifted to the right as
$L_2$ increases, thus confirming that the PDF in the limit $L_2 =
\infty$ exhibits the same power-law decay at any long enough time $t$.

Curiously, the straight part of the curve corresponding to $H(t|r,z)$
lies {\it above} the PDF $H_{\rm 1D}(t|z)$ for the half-line; one can
therefore conclude that the probability of hitting a thin pillar at
time $t$ (large enough) is actually bigger than that for a thick
pillar (with $R_1 = R_2$).  This result sounds counter-intuitive.  To
rationalize it, let us first recall again that the steady-state flux
on a spiky surface is equal to the steady-state flux on an equivalent
absorbing flat surface located at $z = - z_0$, where $z_0 \geq 0$ is
the offset parameter \cite{Grebenkov23}.  As a consequence, the
long-time behavior of the PDF $H(t|r,z)$ can be approximated by that
of $H_{\rm 1D}(t|z+z_0)$ for the half-line with the origin at $-z_0$,
not at $0$.  This is confirmed by the black dotted curve that shows
$H_{\rm 1D}(t|z+z_0)$.  Indeed, this curve lies above $H_{\rm
1D}(t|z)$ at long times thanks to the larger prefactor $z+z_0$.  This
behavior can be rationalized in probabilistic terms.  In fact, any
random trajectory that hits the absorbing point $-z_0$ at time $t$ has
to cross the intermediate level $z = 0$ at an earlier time $t'$.  As
the probability density $H_{\rm 1D}(t|z)$ monotonously decreases at
large $t$, one has $H_{\rm 1D}(t|z+z_0) = H_{\rm 1D}(t'|z) > H_{\rm
1D}(t|z)$.
For comparison, Fig. \ref{fig:Ht_L2}(b) shows the corresponding
survival probability $S(t|r,z)$ and its approximations $S_{\rm
1D}(t|z)$ and $S_{\rm 1D}(t|z+z_0)$, given by Eq. (\ref{eq:St_1D}).

In the following, we assume that $L_2$ is large enough so that the
right tail of the PDF can be approximated by $(z+z_0)/\sqrt{4\pi
Dt^3}$ (with $z > 0$) over a broad range of times.  In this case, the
mean FPT is very large (of the order of $L_2^2/D$) and is thus not
informative.

\subsection{Role of height $L_1$}
\label{sec:L1}

In the previous subsection, we saw how an increase of $L_2$ transforms
an exponential decay of the PDF into a power-law decay.  This is a
direct consequence of the fact that the confining domain $\Omega$
becomes unbounded in the limit $L_2 \to \infty$.
Alternatively, the confining domain $\Omega$ can be made unbounded by
taking the limit $L_1\to \infty$ (with a large but fixed $L_2$).  In
this limit, however, the exponential decay persists even for $L_1 =
\infty$.  In fact, if one formally sets $L_2 \to 0$, the problem
is reduced to diffusion in a semi-infinite tube containing a
semi-infinite absorbing pillar.  Due to the reflecting boundary at $z
= L_2$, this is equivalent to diffusion in an infinite tube with an
infinite pillar, for which diffusion along the tube axis $z$ does not
matter, and the survival probability is determined by diffusion in the
cross-section, i.e., in an annulus between an inner absorbing circle
and an outer reflecting circle.  Despite the fact that the domain is
unbounded, this survival probability admits a spectral expansion
(\ref{eq:St_annulus}) and exhibits an exponential decay at long times.
The decay rate is given by the principal eigenvalue $\lambda_0 =
\alpha_{0,1}^2/R_2^2$, where $\alpha_{0,1}$ is the smallest zero of
Eq. (\ref{eq:alpha1}).  This argument can be extended to any finite
$L_2 > 0$, for which the particle has an additional space $0 < z <
L_2$ for diffusion, so that it is easier to survive and thus
$\lambda_0 \leq \alpha_{0,1}^2/R_2^2$, in agreement with the earlier
established bound (\ref{eq:T_lower}).  At the end of
\ref{sec:long-time}, we provide additional analytic arguments why
there is no power-law decay in the limit $L_1\to\infty$ for any finite
$L_2$.

\begin{figure}
\begin{center}
\includegraphics[width=100mm]{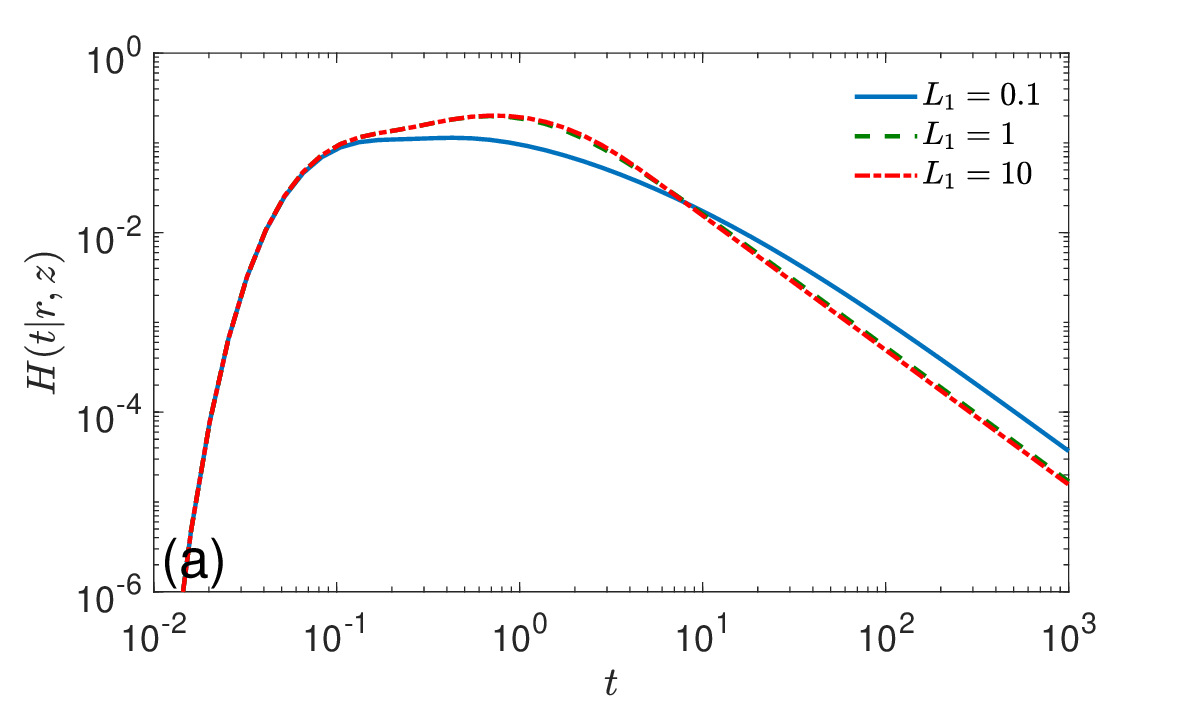} 
\includegraphics[width=100mm]{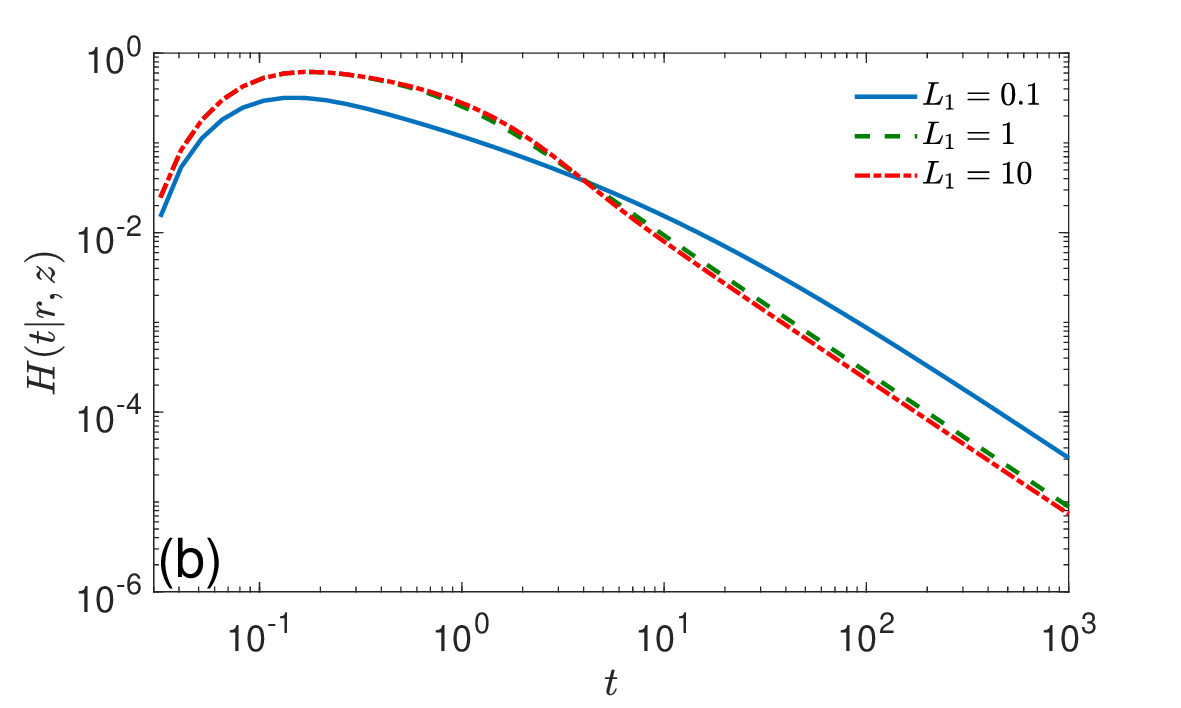} 
\end{center}
\caption{
Probability density $H(t|r,z)$ of the FPT to the absorbing pillar,
with $R_1 = 0.1$, $L_2 = 100$, three values of $L_1$ (see the legend),
and two starting points: $r = 0$, $z = 1$ {\bf (a)} and $r = 1$, $z =
0$ {\bf (b)}. }
\label{fig:Ht_L1}
\end{figure}

Figure \ref{fig:Ht_L1}(a) shows the PDF $H(t|r,z)$ for three values of
$L_1$.  To eliminate the impact of the tube height, we set $L_2 = 100$
and keep the starting point to be above the pillar, with $r = 0$ and
$z = 1$.  One can see that the pillar's height $L_1$ has a low impact
onto the PDF; moreover, the curves for $L_1 = 10$ and $L_1 = 100$ are
almost identical.  This is expected because the matrix $W$ that
determines the coefficients of series representations of
$\tilde{H}(p|r,z)$, depends on $L_1$ only through the elements
$B^{(1)}_n$ given by Eq. (\ref{eq:Bn1}), in which $h_1 = L_1/R_2$
enters in the argument of $\ctanh(\alpha_{n,1}' h_1)$, with
$\alpha_{n,1}'$ given by Eq. (\ref{eq:alpha1p}).  When $\alpha_{n,1}'
h_1 \geq \alpha_{0,1}' h_1 \gg 1$, the elements $B^{(1)}_n$ do not
almost depend on $h_1$, implying the independence of
$\tilde{H}(p|r,z)$ and thus of $H(t|r,z)$ on the height $L_1$, when
$L_1$ is large enough, in agreement with panel (a).  This argument is
valid for any $z > 0$, i.e., when the particle starts above the
pillar.

In turn, if the particle starts on a side of the pillar ($z < 0$), the
dependence on $L_1$ is stronger because $L_1$ also appears in the
function $s_{n,1}(z)$ given by Eq. (\ref{eq:s1}).  Panel (b) of
Fig. \ref{fig:Ht_L1} illustrates this effect for the starting point at
$(r,z) = (1,0)$, i.e., at the outer reflecting boundary on the level
of the pillar's top.  Even here, the effect of $L_1$ is moderate,
especially for large $L_1$.  In the next subsection, we inspect the
dependence on the height of the starting point in the case of long
enough pillars.

It is worth noting that the opposite limit $L_1 \to 0$ corresponds to
a periodic array of absorbing disks on the reflecting plane.
Steady-state diffusion towards such configurations was studied earlier
(see \cite{Berezhkovskii_2004,Bernoff18,Bernoff18b,Grebenkov23} and
references therein).  For any small but strictly positive $L_1$, the
elements $B^{(1)}_n$ behave as $R_2/(\alpha'_{n,1})^2/L_1$ for $n \ll
n_0$, and as $1/\alpha'_{n,1}$ for $n \gg n_0$, where the index $n_0$
is determined by the condition $\alpha'_{n_0,1} \sim R_2/L_1$.  As a
consequence, the elements with moderate $n$ are getting larger as
$L_1\to 0$, but the asymptotic form of this matrix remains unchanged.
One sees that the analysis of the limit $L_1\to 0$ is much more subtle
and is beyond the scope of this paper.

\subsection{Role of position $z$}

To analyze the role of the vertical position $z$ of the starting
point, we fix the pillar's height $L_1 = 10$ and keep again $L_2 =
100$.

\begin{figure}
\begin{center}
\includegraphics[width=100mm]{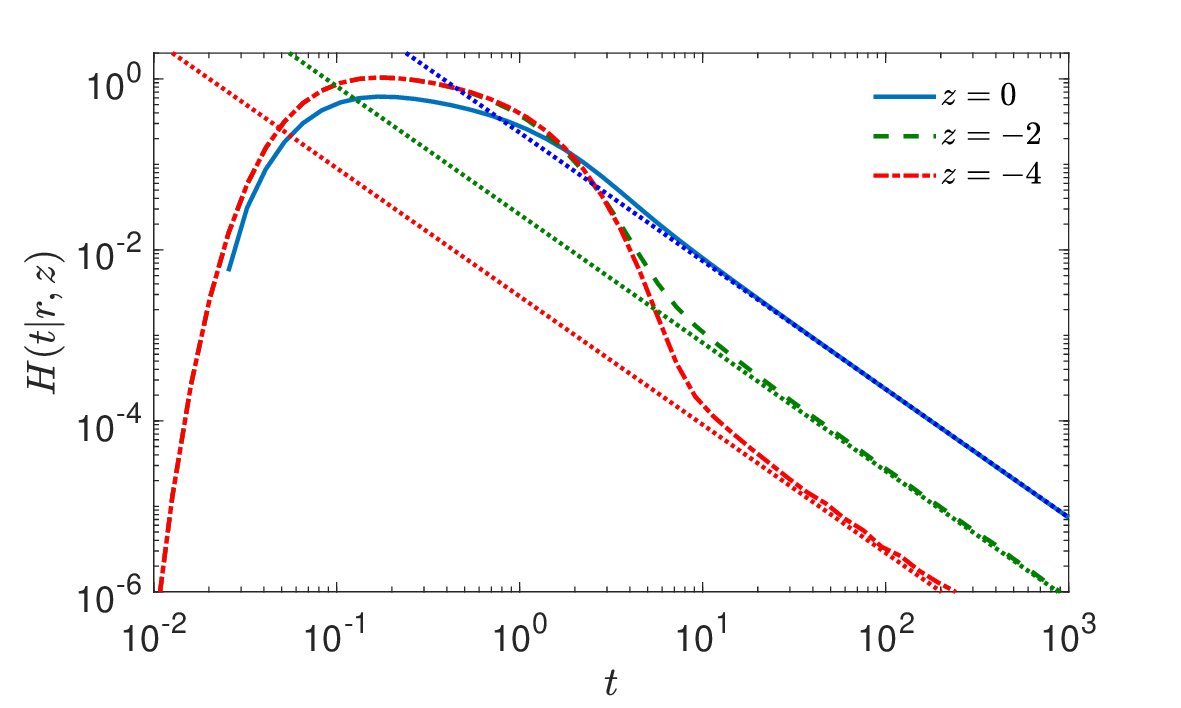} 
\end{center}
\caption{
Probability density $H(t|r,z)$ of the FPT to the absorbing pillar,
with $R_1 = 0.1$, $L_1 = 10$, $L_2 = 100$, and different locations of
the starting point $(r,z)$, with $r = 1$ and three values of $z$ (see
the legend).  Three dotted lines present the long-time behavior
(\ref{eq:Ht_long}), with $\alpha_{0,1}\approx 1.10$, $C(1) \approx
1.11$, and $z_0 \approx 0.75$ found in \cite{Grebenkov23}.  Note that
the solid blue is not shown at times $t \lesssim 0.03$ due to
numerical instabilities in the inversion of the Laplace transform.}
\label{fig:Ht_z}
\end{figure}

Figure \ref{fig:Ht_z} shows the PDF $H(t|r,z)$ evaluated at $r = 1$
(i.e., at the outer cylindrical boundary) and three values of $z$:
$0$, $-2$, and $-4$.  Expectedly, the short-time behavior, which is
determined by ``direct'' trajectories and thus by the distance to the
absorbing pillar, is almost identical for three cases.  The long-time
behavior exhibits the same power-law decay $t^{-3/2}$ but with
different prefactors (we recall that the exponential cut-off due to
the boundness of the domain appears at much longer times exceeding
$L_2^2/D = 10^4$).  When $z = 0$, one can still rely on the
one-dimensional PDF $H_{\rm 1D}(t|z_0)$ from Eq. (\ref{eq:Ht_1D}) with
the offset parameter $z_0$ accounting for the reduced radius $R_1$ of
the pillar (as compared to $R_2$).  The resulting long-time asymptotic
behavior $z_0/\sqrt{4\pi Dt^3}$, which is shown by blue dotted line,
is in excellent agreement with $H(t|r,0)$.  

In order to characterize the reduced amplitude of this line for
negative $z$, we employ the following argument.  When the particle
starts in the region $z < 0$, one can split random trajectories in two
groups: (i) those that arrived onto the pillar without crossing the
level $z = 0$, and (ii) those that crossed the level $z = 0$.  For the
first group, the survival probability decays exponentially in time,
with the decay time of the order of $T_1 = R_2^2 /(\alpha_{0,1}^2 D)$
(see Sec. \ref{sec:L1}).  At times $t \gg T_1$, this contribution is
negligible, and the long-time asymptotic behavior is mainly determined
by the trajectories of the second group that managed to escape from
the region with $z < 0$ and thus can explore the elongated upper
region with $z > 0$.  In a first approximation, the long-time behavior
of $H(t|r,z)$ can thus be approximated again by $z_0/\sqrt{4\pi
Dt^3}$, multiplied by the fraction of trajectories in the second
group.  This fraction is given by the splitting probability computed
in \ref{sec:splitting}.  When $|z|/R_2$ is large enough, the splitting
probability can be approximated by the leading term, see
Eq. (\ref{eq:u_c0}), so that
\begin{equation}   \label{eq:Ht_long}
H(t|r,z) \approx C(r)\, e^{\alpha_{0,1} z/R_2} \frac{z_0}{\sqrt{4\pi Dt^3}} \,,
\end{equation}
where the amplitude $C(r)$ is defined by Eq. (\ref{eq:Cr_splitting}).
The good accuracy of this asymptotic relation is confirmed on
Fig. \ref{fig:Ht_z}.  

\subsection{Role of radius $R_1$}
\label{sec:R1}

We analyze the role of the pillar's radius $R_1$.  When $R_1 = R_2$,
the pillar fills the tube, there is no diffusion in the region $z <
0$, while the survival probability for the upper region $z > 0$ is
simply given by $S_{\rm 1D}(t|z)$ for diffusion on the interval
$(0,L_2)$, see Eq. (\ref{eq:St_interval}).  When $R_1$ is smaller but
still comparable to $R_2$, the particle that managed to enter the
region $z < 0$, is rapidly absorbed by the side surface of the pillar.
In this light, configurations with long but thin pillar (i.e., $R_1
\ll R_2$) seem to be most interesting from both theoretical and
practical points of view.

In the limit $\rho = R_1/R_2\to 0$, the pillar shrinks to a needle,
i.e., a finite segment or a half-line, which are ``invisible'' for
Brownian motion \cite{Morters}.  In other words, an infinitely thin
pillar cannot absorb the particle, and the survival probability is
equal to $1$ in this limit.  However, the approach to this limit is
very slow.  As discussed in \cite{Grebenkov23}, the asymptotic
behavior of Bessel functions implies that
\begin{equation}  \label{eq:alpha0_rho0}
\alpha_{0,1} \approx \frac{\sqrt{2}}{\sqrt{\ln(1/\rho) - 3/4}}  \quad (\rho \to 0).
\end{equation}
One sees that $\alpha_{0,1}$ indeed vanishes as $\rho \to 0$ but
extremely slowly.  In particular, this slow decay ensures that the
decay time $R_2^2/(\alpha_{0,1}^2 D)$ associated to planar diffusion
is generally (much) smaller than the decay time $4L_2^2/(\pi^2 D)$
associated to diffusion in the upper region when $L_2/R_2 \gg 1$.  In
fact, this occurs when
\begin{equation}
\frac{L_2}{R_2} > \frac{\pi}{2\sqrt{2}} \sqrt{\ln(1/\rho) - 3/4} \,.
\end{equation}
For instance, if $\rho = 10^{-2}$, this inequality leads to a moderate
constraint $L_2/R_2 > 2.18$.  Alternatively, one can get a bound on
the relative radius of the pillar:
\begin{equation}
\rho > \exp\bigl(-3/4 - (8/\pi^2)(L_2/R_2)^2\bigr) .  
\end{equation}
Even for a moderate value $L_2/R_2 = 5$, the decay time associated to
one-dimensional diffusion is dominant whenever the relative radius
exceeds $7.5\cdot 10^{-10}$, i.e., in any relevant setting.

\subsection{Role of proximity to the pillar}

In previous sections, the starting point was located relatively far
from the pillar, with the distance to the pillar being equal to $R_2$.
Let us now look at the effect of proximity of the starting point to
the pillar.

If the particle is released from a point $(r,z)$ near the top of the
pillar (i.e., $0 < z \ll R_1$ and $r \ll R_1$), the particle explores
at short times the vicinity of a flat boundary, as it was near an
absorbing plane in the upper half-space.  As a consequence, the PDF of
the FPT is accurately described by $H_{\rm 1D}(t|z)$ from
Eq. (\ref{eq:Ht_1D}).  As time increases, the particle starts to
``feel'' that the top of the pillar has a finite radius, and thus
deviates from Eq. (\ref{eq:Ht_1D}).  Note that if $L_2$ is large
enough, the long-time behavior is again one-dimensional and given by
$H_{\rm 1D}(t|z+z_0)$, which exhibits the same long-time $t^{-3/2}$
behavior but with a higher amplitude (see Sec. \ref{sec:L2}).

Let us now examine another setting when the particle is released from
a point $(r,z)$ near the side of the pillar (i.e., $R_1 < r \ll R_2$
and $z < 0$ with $|z| \gg R_2$).  At short times, the particle
explores a vicinity of the curved surface of a cylindrical pillar of
radius $R_1$, as if it diffused outside an infinite absorbing cylinder
of radius $R_1$ in the three-dimensional space.  This is equivalent to
planar diffusion outside an absorbing circle of radius $R_1$ (at this
time, diffusion along the $z$ axis does not matter yet).  In this
case, the survival probability is known to exhibit a very slow decay
(see, e.g., \cite{Koplik94,Koplik95,Levitz08})
\begin{equation}  \label{eq:S_2D_asympt0}
S_{\rm 2D}(t|r) \approx \frac{2 \ln(r/R_1)}{\ln (Dt/R_1^2)}  \quad (t\to\infty),
\end{equation}
from which
\begin{equation}  \label{eq:H_2D_asympt0}
H_{\rm 2D}(t|r) \approx \frac{2 \ln(r/R_1)}{t[\ln (Dt/R_1^2)]^2}  \quad (t\to\infty).
\end{equation}
A more accurate expression for the asymptotic behavior of the PDF was
given in \cite{Grebenkov18d,Grebenkov21}
\begin{equation}  \label{eq:Ht_2D_asympt}
H_{\rm 2D}(t|r) \approx \frac{2 \ln(r/R_1)}{t \bigl[\pi^2 + \bigl(\ln (R_1^2/(4Dt)) + 2\gamma\bigr)^2\bigr]} \,,
\end{equation}
where $\gamma \approx 0.5772$ is the Euler constant.  
Note that the integral of this expression yields
\begin{equation}  \label{eq:St_2D_asympt}
S_{\rm 2D}(t|r) \approx \frac{2 \ln(r/R_1)}{\pi}  \arctan\biggl(\frac{\pi}{\ln (4Dt/R_1^2) - 2\gamma}\biggr).
\end{equation}
At very large $t$, one retrieves Eqs. (\ref{eq:S_2D_asympt0},
\ref{eq:H_2D_asympt0}).

These expressions provide the long-time asymptotic behavior for planar
diffusion outside an absorbing circle.  In our case, however, these
expressions yield the transient behavior at {\it intermediate} time
scales, until the particle starts to ``feel'' the confinement.  As
time increases further, the motion of the particle is affected by
confinement, and the asymptotic behavior changes.  This change occurs
at the time needed to reach the outer boundary of radius $R_2$.  The
latter can be estimated as the decay time $T_2 = R_2^2/(j_{0,1}^2D)
\approx 0.17$ of the survival probability of a particle diffusing
inside a disk of radius $R_2$ with the absorbing boundary, where
$j_{0,1} \approx 2.4048$ is the first positive zero of the Bessel
function $J_0(z)$.  At even longer times, the particle may reach the
upper region (with $z > 0$) and diffuse further away from the pillar.
If $L_2$ is large enough, another intermediate regime with the
$t^{-3/2}$ decay is established, as discussed in Sec. \ref{sec:L2}.
This regime is terminated by an exponential cut-off with the decay
time $T$ discussed in Sec. \ref{sec:decay-time}.

\begin{figure}
\begin{center}
\includegraphics[width=100mm]{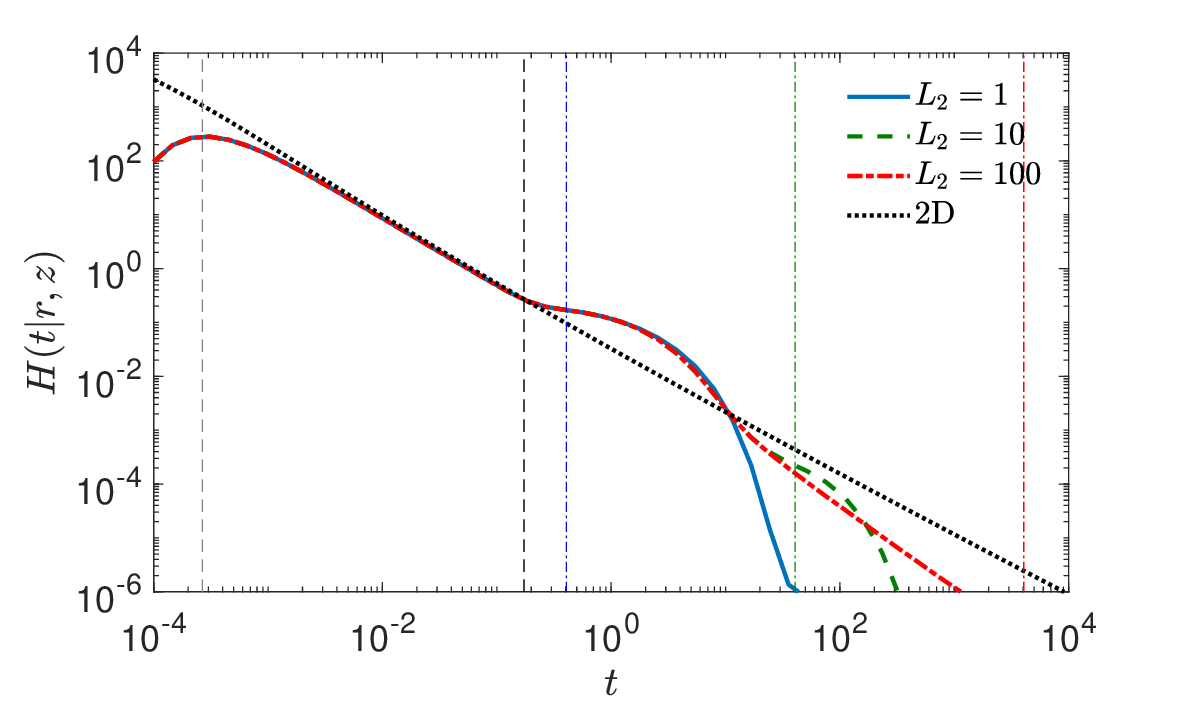} 
\end{center}
\caption{
Probability density $H(t|r,z)$ of the FPT to the absorbing pillar,
with $R_1 = 0.01$, $L_1 = 10$, $(r,z) = (0.05,-2)$, and three values
of $L_2$ (see the legend).  Dotted line presents the long-time
behavior (\ref{eq:Ht_2D_asympt}) for planar diffusion.  Vertical
dashed lines indicate several time scales: $(r-R_1)^2/(6D) \approx
2.7\cdot 10^{-4}$ is the most probable FPT, $R_2^2/(j_{0,1}^2 D)
\approx 0.17$ is the decay time for planar diffusion towards an
absorbing disk of radius $R_2$, where $j_{0,1} \approx 2.4048$ is the
first positive zero of $J_0(z)$, while $4\cdot 1^2/(\pi^2D) \approx
0.4$, $4\cdot 10^2/(\pi^2D) \approx 40$ and $4\cdot 100^2/(\pi^2D)
\approx 4000$ are the decay times for one-dimensional diffusion with
$L_2 = 1, 10$ and $100$, respectively.}
\label{fig:Ht_r}
\end{figure}

Figure \ref{fig:Ht_r} illustrates the effect of proximity of the
starting point to the pillar's side.  Here we consider a thin pillar
($R_1 = 0.01$) and locate the starting point $(r,z)$ close to the
pillar's boundary at $r = 0.05$ and $z = -2$.  At short times, at
which the particle does not ``feel'' the presence of the outer
reflecting boundary, one retrieves the asymptotic behavior
(\ref{eq:Ht_2D_asympt}) reminiscent of planar diffusion.  At the time
scale $T_2$ shown by a black dashed vertical line, there appear
deviations from Eq. (\ref{eq:Ht_2D_asympt}).  At even longer times,
one observes another intermediate regime with the $t^{-3/2}$ decay,
which corresponds to a diffusive exploration of the upper region (with
$z > 0$).  This regime is terminated by an exponential cut-off at the
time decay $4L_2^2/(\pi^2 D)$ for one-dimensional diffusion, which is
equal to $0.4$, $40$ and $4000$ at $L_2 = 1, 10$, and $100$,
respectively.  Clearly, this time scale for $L_2 = 1$ is too close to
$T_2 \approx 0.17$ so that the intermediate regime does not exist.  In
turn, it is clearly visible at $L_2 = 100$.

\subsection{Comparison with the self-consistent approximation}

In Ref. \cite{Grebenkov18d}, the distribution of the FPT was studied
for a similar configuration of two coaxial cylinders of radii $R_1$
and $R_2$, capped by the parallel planes at $z = -L_1$ and $z = L_2$.
The absorbing region was located on a lower part ($-L_1 < z < 0$) of
the inner cylinder, while its upper part ($0 < z < L_2$) was
reflecting.  While this configuration resembles our setting with an
absorbing pillar, there is a significant difference: the upper inner
cylinder was impenetrable to diffusing particles so that the top of
the pillar was inaccessible in \cite{Grebenkov18d}.  When the inner
cylinder is very thin, such a difference does not seem to be
significant.  In contrast, if the inner cylinder is moderately thin,
the excluded volume may play an important role.  In particular, the
top of the pillar may have very high chances to absorb the particle
arriving from a remote point above the pillar, thus screening the side
of the pillar.  Note also that the limit $L_1 \to 0$ is totally
different in two settings: in our case, the pillar shrinks to a disk,
which can still absorb particles; in turn, in the setting of
Ref. \cite{Grebenkov18d}, the absorbing region was exclusively located
on the side of the pillar, and the survival probability becomes equal
to $1$ in the limit $L_1\to 0$.  We conclude that our study provides
complementary insights onto diffusion-controlled reactions in such
domains.

\section{Conclusion}
\label{sec:conclusion}

In this paper, we investigated the distribution of the FPT to a
periodic array of absorbing pillars confined between two parallel
reflecting planes.  The replacement of a periodic cell of the original
system by a cylindrical tube with reflecting boundary that englobes a
single pillar allowed us to solve exactly the modified Helmholtz
equation in cylindrical coordinates.  For this purpose, we adopted the
mode matching method that we recently developed for studying
steady-state diffusion governed by the Laplace equation
\cite{Grebenkov23}.  In this way, we managed to obtain an exact
representation of the Laplace-transformed PDF $\tilde{H}(p|r,z)$ of
the FPT.  Despite the need for a numerical inversion of a truncated
matrix with explicitly known elements, this solution presents many
advantages: (i) analytical dependence of $\tilde{H}(p|r,z)$ on the
starting point $(r,z)$; (ii) rapid convergence and therefore very fast
numerical computation; (iii) identification of respective roles of
different geometric parameters onto the solution; and (iv) asymptotic
analysis.  In particular, the method was fast enough to undertake an
inverse Laplace transform numerically and to get the survival
probability $S(t|r,z)$ and the PDF $H(t|r,z)$ in time domain.

From a theoretical point of view, an absorbing pillar surrounded by a
reflecting boundary is a rich geometric model to investigate various
aspects of the FPT distribution.  In fact, former theoretical studies
were focused on simpler geometric settings like coaxial cylinders or
concentric spheres.  In turn, the current model has four geometric
parameters: the pillar's height $L_1$ and radius $R_1$, the distance
$L_2$ to the upper plane, and the radius $R_2$ of the outer reflecting
boundary (which is also related to the inter-pillar distance in the
original periodic array of pillars).  As a consequence, different
asymptotic regimes can emerge and even co-exist.  For instance,
Fig. \ref{fig:Ht_r} presented the PDF with four distinct regimes: (i)
a universal short-time behavior $e^{-(r-R_1)^2/(4Dt)}/t^{3/2}$
governed by ``direct trajectories'' (left tail), an intermediate
behavior $1/(t \ln^2 (Dt/R_1^2))$ due to effectively planar diffusion,
an intermediate behavior $t^{-3/2}$ due to effectively one-dimensional
exploration of the upper region, and a universal exponential cut-off
$e^{-t/T}$ due to confinement.  Even though each of these regimes have
been studied in the past, we are unaware of earlier observations of
all these features in a single PDF.  In order to better understand
these features, we discussed how different geometric parameters affect
the distribution.

From a practical point of view, spiky coatings have recently drown
significant attention due to the rapid progress in fabrication
technology and favorable performance in many applications such as
superhydrophobic materials \cite{Davis_2010}, filtration
\cite{Ramon_2012,Ramon_2013}, sensing systems
\cite{Nair_2007,Chen_2022}, selective protein separation
\cite{Borberg19}, to name but a few.  At the same time, a theoretical
description of their trapping efficiency was still missing, especially
in a transient time-dependent regime.  To our knowledge, this is the
first study of the FPT in such structures.  We stress that the derived
exact solution goes far beyond the conventional mean FPT, which is
uninformative and actually misleading if the upper plane is located
far away from the pillars.  We therefore expect that the presented
method and solution may guide experimentalists in the intelligent
design of spiky coatings with desired trapping properties.

\section*{Acknowledgments}

D.S.G. acknowledges the Alexander von Humboldt Foundation for support
within a Bessel Prize award.  A.T.S. thanks Paul A. Martin for many
illuminating discussions.

\appendix
\section{Exact solution}
\label{sec:exact}

In this Appendix, we provide the details of the derivation of the
exact solution of Eqs. (\ref{eq:problem_u}).  This derivation closely
follows the Appendix of Ref. \cite{Grebenkov23}, in which the mode
matching method was used to solve the Laplace equation.  Even though
many notations and equations are identical, we reproduce the whole
derivation to highlight subtle modifications that are required for
solving Eqs. (\ref{eq:problem_u}).

\subsection{Derivation of the solution}
\label{sec:derivation}

Due to the axial symmetry, the boundary value problem
(\ref{eq:problem_u}) is actually a two-dimensional problem in an
L-shape region (see Fig. \ref{fig:scheme}(c)).  Note that one has to
add the Neumann boundary condition,
\begin{equation}  \label{eq:cond_r0}
\partial_r \tilde{H} = 0 \quad (r = 0,~ 0 < z < L_2),
\end{equation} 
to account for the regularity and axial symmetry of the problem.  One
can search for its solution separately in two rectangular subdomains,
$\Omega_1 = (R_1,R_2) \times (-L_1,0)$ and $\Omega_2 = (0,R_2) \times
(0,L_2)$, and then match them at the junction interval (at $z = 0$).

A general solution in $\Omega_1$ reads 
\begin{equation}   \label{eq:u1}
\tilde{H}(p|r,z) = w(r/R_2) - \sum\limits_{n=0}^\infty c_{n,1} \, v_{n,1}(r/R_2) \, s_{n,1}(z) \,,
\end{equation}
with unknown coefficients $c_{n,1}$, where
\begin{equation}  \label{eq:u01}
w(\r) = \frac{K_1(\alpha) I_0(\alpha \r) + I_1(\alpha) K_0(\alpha \r)}
{K_1(\alpha) I_0(\alpha \rho) + I_1(\alpha) K_0(\alpha \rho)} \,,
\end{equation}
with $\alpha = R_2 \sqrt{p/D}$, $\rho = R_1/R_2$, $\r$ denoting
dimensionless radius,
\begin{equation}  
s_{n,1}(z) = \frac{\cosh(\alpha_{n,1}' (L_1+z)/R_2)}{\cosh(\alpha_{n,1}' L_1/R_2)}   \label{eq:s1}
\end{equation}
and
\begin{equation}  \label{eq:vn1}
v_{n,1}(\r) = e_n \, w_n(\r) ,
\end{equation}
with
\begin{equation}
w_n(\r) = J_1(\alpha_{n,1}) Y_0(\alpha_{n,1} \r) - Y_1(\alpha_{n,1}) J_0(\alpha_{n,1} \r),
\end{equation}
and we used $J'_0(z) = -J_1(z)$, $Y'_0(z) = -Y_1(z)$, prime denotes
the derivative, $J_\nu(z)$ and $Y_\nu(z)$ are the Bessel functions of
the first and second kind, respectively, and $I_\nu(z)$ and $K_\nu(z)$
are the modified Bessel functions.  The prefactor
\begin{equation}
e_n = \frac{\sqrt{2}}{\sqrt{[w_n(1)]^2 - \rho^2 [w'_n(\rho)/\alpha_{n,1}]^2}}
\end{equation}
ensures the normalization of the radial function $v_{n,1}(\r)$:
\begin{equation}
\int\limits_\rho^1 d\r \, \r \, [v_{n,1}(\r)]^2 = 1 ,
\end{equation}
where we used
\begin{align*}
\int\limits_\rho^1 d\r \, \r\, w_n^2(\r) & = \frac{1}{2\alpha_{n,1}^2} 
\biggl(\r^2 [w'_n(\r)]^2 + \alpha_{n,1}^2 \r^2 [w_n(\r)]^2\biggr)_\rho^1 \\ & = \frac{[w_n(1)]^2 - \rho^2 [w'_n(\rho)/\alpha_{n,1}]^2}{2} ,
\end{align*}  
with $w_n(\rho) = 0$ and $w'_n(1) = 0$ being employed.
By construction, $\tilde{H}(p|r,z)$ from Eq. (\ref{eq:u1}) satisfies
Eqs. (\ref{eq:Helm}, \ref{eq:cond_outer}, \ref{eq:cond_bottom}).  The
parameters $\alpha_{n,1}$ are obtained by imposing the condition
(\ref{eq:cond_inner}) at $r = R_1$ (i.e., setting $w_n(\rho) = 0$) and
solving the resulting equation
\begin{equation}  \label{eq:alpha1}
Y_1(\alpha_{n,1}) J_0(\alpha_{n,1} \rho) - J_1(\alpha_{n,1}) Y_0(\alpha_{n,1} \rho) = 0  .
\end{equation}
This equation has infinitely many positive solutions
$\{\alpha_{n,1}\}$, which are enumerated by $n=0,1,2,\ldots$ in an
increasing order \cite{Watson}.  As $\{v_{n,1}(\r)\}$ are the
eigenfunctions of the differential operator $\partial_r^2 + (1/r)
\partial_r$, they form a complete orthonormal basis in the space
$L_2(\rho,1)$ of $r$-weighted square-integrable functions on
$(\rho,1)$.  Finally, one sets
\begin{equation}  \label{eq:alpha1p}
\alpha_{n,1}' = \sqrt{\alpha_{n,1}^2 + R_2^2 p/D} .
\end{equation}

A general solution in $\Omega_2$ reads 
\begin{equation}  \label{eq:u2}
\tilde{H}(p|r,z) = \sum\limits_{n=0}^\infty c_{n,2}\, v_{n,2}(r/R_2) \, s_{n,2}(z) ,
\end{equation}
with unknown coefficients $c_{n,2}$, where
\begin{equation}  \label{eq:v2}
v_{n,2}(\r) =  \frac{J_0(\alpha_{n,2}\r)}{J_0(\alpha_{n,2})} \, ,
\end{equation} 
and
\begin{equation} \label{eq:s2}
s_{n,2}(z) = \frac{\cosh(\alpha'_{n,2} (L_2-z)/R_2)}{\cosh(\alpha'_{n,2} L_2/R_2)} \,. 
\end{equation}
By construction, $\tilde{H}(p|r,z)$ from Eq. (\ref{eq:u2}) satisfies
Eqs. (\ref{eq:Helm}, \ref{eq:cond_top}, \ref{eq:cond_r0}).  The
parameters $\{\alpha_{n,2}\}$ are obtained by imposing the condition
(\ref{eq:cond_outer}), which reads as
\begin{equation}  \label{eq:alpha_n2}
J_1(\alpha_{n,2}) = 0  \quad (n=0,1,2,\ldots).
\end{equation}
This equation has infinitely many positive solutions $\{
\alpha_{n,2}\}$, which are enumerated by $n = 0,1,2,\ldots$ in an
increasing order \cite{Watson}.  The prefactor in Eq. (\ref{eq:v2})
ensures the normalization:
\begin{equation}  \label{eq:v2norm}
\int\limits_0^1 d\r \, \r \, [v_{n,2}(\r)]^2 = \frac{1}{2} \,.
\end{equation}
As $\{\sqrt{2} \, v_{n,2}(\r)\}$ are the eigenfunctions of the
differential operator $\partial_r^2 + (1/r) \partial_r$, they form a
complete orthonormal basis in the space $L_2(0,1)$.  Finally, one sets
\begin{equation}  \label{eq:alpha2p}
\alpha_{n,2}' = \sqrt{\alpha_{n,2}^2 + R_2^2 p/D} .  
\end{equation}
Note that $\alpha_{0,2} = 0$ so that $\alpha_{0,2}' = \alpha$.

The unknown coefficients $c_{n,1}$ and $c_{n,2}$ are then determined
by matching the representations (\ref{eq:u1}, \ref{eq:u2}) at $z = 0$,
i.e., by requiring the continuity of $\tilde{H}(p|r,z)$ and of its
derivative $\partial_z \tilde{H}(p|r,z)$.  The second condition, which
should be satisfied for any $R_1 < r < R_2$, reads
\begin{align}  \label{eq:matching2}
& R_2 (\partial_z \tilde{H})_{z=0^{-}} = \sum\limits_{n=0}^\infty \tilde{c}_{n,1} v_{n,1}(r/R_2) \,   
 = \sum\limits_{n=0}^\infty \tilde{c}_{n,2} v_{n,2}(r/R_2) = R_2 (\partial_z \tilde{H})_{z=0^+} ,
\end{align}
where
\begin{equation}
\tilde{c}_{n,1} = \frac{c_{n,1}}{B_n^{(1)}}  \,, \qquad 
\tilde{c}_{n,2} = - c_{n,2} B_n^{(2)} ,  
\end{equation}
and
\begin{align}  \label{eq:Bn1}
B_n^{(1)} & = \frac{1}{R_2 s'_{n,1}(0)} = \frac{\ctanh(\alpha_{n,1}' L_1/R_2)}{\alpha_{n,1}'} \,, \\  \label{eq:Bn2}
B_n^{(2)} & = - R_2 s'_{n,2}(0) = \alpha'_{n,2}\, \tanh(\alpha'_{n,2} L_2/R_2) ,
\end{align}
with $\tanh(z)$ and $\ctanh(z)$ denoting the hyperbolic tangent and
cotangent functions, respectively.  Multiplying
Eq. (\ref{eq:matching2}) by $\r\, v_{k,1}(\r)$ and integrating from
$\rho$ to $1$, one gets
\begin{align*}
& \sum\limits_{n=0}^\infty \tilde{c}_{n,2} \int\limits_\beta^1 d\r \, \r \, v_{k,1}(\r) \,  v_{n,2}(\r) 
 = \tilde{c}_{k,1} 
\end{align*}
due to orthogonality of $\{v_{k,1}(\r)\}$.  Setting
\begin{equation}   \label{eq:A_def}
A_{k,n} = \int\limits_\rho^1 d\r \, \r \, v_{k,1}(\r) \,  v_{n,2}(\r) , 
\end{equation}
we can rewrite the above equations as
\begin{equation}  \label{eq:ck1}
c_{k,1} = B_k^{(1)} \sum\limits_{n=0}^\infty A_{k,n} B_n^{(2)} c_{n,2} \,.
\end{equation} 
Moreover, as the radial functions $v_{k,1}(\r)$ and $v_{k,2}(\r)$ are
linear combinations of Bessel functions of the {\it same order}, the
integral in Eq. (\ref{eq:A_def}) can be found explicitly:
\begin{align}  \label{eq:A}
A_{k,n} & = \biggl(\r\frac{v_{k,1}(\r) v'_{n,2}(\r) - v'_{k,1}(\r) v_{n,2}(\r)}{\alpha_{k,1}^2 - \alpha_{n,2}^2}\biggr)_{\r = \rho}^{1}   
 = \frac{\rho\, v'_{k,1}(\rho)\, v_{n,2}(\rho)}{\alpha_{k,1}^2 - \alpha_{n,2}^2} \,,
\end{align}
where we used the boundary conditions $v_{k,1}(\rho) = v'_{k,1}(1) =
v'_{n,2}(1) = 0$.

Similarly, we impose the continuity of the function $\tilde{H}(p|r,z)$
at $z=0$, together with Eq. (\ref{eq:cond_disk}):
\begin{equation}  \label{eq:u_continuity}
\tilde{H}(p|r,0^+) = \begin{cases} 1  \hskip 25mm (0 < r < R_1), \cr   \tilde{H}(p|r,0^-) \qquad (R_1 < r < R_2). \end{cases}
\end{equation}
Multiplying this relation by $\r v_{k,2}(\r)$ and integrating from
$0$ to $1$, we get
\begin{equation}  \label{eq:auxil11}
\frac{c_{k,2}}{2} = - \frac{\rho v'_{k,2}(\rho)}{\alpha_{k,2}^2} 
+ \int\limits_{\rho}^1 d\r \, \r \, v_{k,2}(\r) \, w(\r) - \sum\limits_{n=0}^\infty c_{n,1} \, A_{n,k},
\end{equation}
where we used the orthogonality of functions $\{v_{k,2}(\r)\}$ and
their normalization (\ref{eq:v2norm}); note that the first term is
equal to $\rho^2/2$ for $k = 0$.  Substituting $c_{n,1}$ from
Eq. (\ref{eq:ck1}), we get
\begin{equation} 
c_{k,2} + 2\sum\limits_{n=0}^\infty A_{n,k} B_n^{(1)} \sum\limits_{n'=0}^\infty A_{n,n'} B_{n'}^{(2)} c_{n',2} = V_k \,,
\end{equation}
where
\begin{equation}
\frac{V_k}{2} = - \frac{\rho v'_{k,2}(\rho)}{\alpha_{k,2}^2} + \int\limits_{\rho}^1 d\r \, \r \, v_{k,2}(\r) w(\r) .
\end{equation}
In analogy to Eq. (\ref{eq:A}), one can compute the second integral explicitly:
\begin{align}  \nonumber
\frac{V_k}{2} & = - \frac{\rho v'_{k,2}(\rho)}{\alpha_{k,2}^2} + 
\biggl(\r\frac{v_{k,2}(\r) w'(\r) - v'_{k,2}(\r) w(\r)}{\alpha_{k,2}^2 + \alpha^2}\biggr)_{\r = \rho}^{1} \\  \label{eq:V}
& = - \frac{\rho v'_{k,2}(\rho)}{\alpha_{k,2}^2} - \rho \frac{v_{k,2}(\rho)\, w'(\rho) - v'_{k,2}(\rho)}{\alpha_{k,2}^2 + \alpha^2} \,,
\end{align}  
where we used that $w(\rho) = 1$.  It is convenient to re-arrange
two sums in Eq. (\ref{eq:auxil11}) as
\begin{equation} \label{eq:system}
c_{k,2} + \sum\limits_{n=0}^\infty W_{k,n}\, c_{n,2}  = V_k  \quad (k = 0,1,2,\ldots),
\end{equation}
where
\begin{equation}  \label{eq:Wdef}
W_{k,n} = 2\sum\limits_{n'=0}^\infty A_{n',k} B_{n'}^{(1)}  A_{n',n} B_{n}^{(2)} \,,
\end{equation}
i.e., we got the infinite system of linear algebraic equations for the
unknown coefficients $c_{k,2}$ with $k=0,1,2,\ldots$.  To compute
these coefficients, one needs to construct the infinite-dimensional
matrix $W$ and then to invert the matrix $I + W$, where $I$ is the
identity matrix.  In practice, one can truncate the matrix $I + W$ to
a finite size $N\times N$ and then perform the inversion numerically.
Once the coefficients $c_{n,2}$ are found, one can determine $c_{n,1}$
according to Eq. (\ref{eq:ck1}).  This completes the construction of
the exact solution of the problem (\ref{eq:problem_u}).  Even though
this construction involves numerical inversion of the truncated
matrix, the obtained expressions (\ref{eq:u1}, \ref{eq:u2}) provides
an explicit analytical dependence of $\tilde{H}(p|r,z)$ on $r$ and $z$
via the functions $v_n(r/R_2)$ and $s_n(z)$.  Moreover, the accuracy
of the numerical computation of $\tilde{H}(p|r,z)$ rapidly improves as
the truncation order $N$ increases.  In most cases, one can use
moderate values of $N$ (say, few tens) to get very accurate results.

Importantly, the structure of the exact solution reveals how different
geometric parameters can affect the FPT distribution: the pillar
height $L_1$ enters {\it only} via $B_n^{(1)}$, the distance to the
source $L_2$ enters {\it only} via $B_n^{(2)}$, so that the matrix $A$
does not depend on $L_1$ and $L_2$.  Similarly, the matrix $A$ does
not depend on $p$.  These properties can be used for deriving various
asymptotic behaviors (see, e.g., \ref{sec:long-time}).  For instance,
in the limit $p\to 0$, one has $w(\r) \to 1$ and $w'(\rho)
\approx (\rho - 1/\rho)\alpha^2/2$.  As a consequence, one gets
\begin{equation}
\frac{V_k}{2} \approx - \rho \frac{v_{k,2}(\rho) w'(\rho)}{\alpha_{k,2}^2} \to 0  \qquad (k > 0) ,
\end{equation}
whereas $V_0 \to 1$.  Moreover, one has $B_0^{(2)} \to 0$ so that
$W_{k,0} \to 0$.  In this limit, one deals with the homogeneous system
of linear equations,
\begin{equation}
c_{k,2} + \sum\limits_{n=1}^\infty W_{k,n} c_{n,2} = 0  \qquad (k=1,2,\ldots),
\end{equation}
which has the trivial solution $c_{k,2} = 0$ for all $k > 0$.  In
addition, one gets $c_{0,2} = 1$ and therefore retrieves the expected
normalization: 
\begin{equation}
\int\limits_0^\infty dt \, H(t|r,z) = \tilde{H}(0|r,z) = 1.
\end{equation}

\subsection{Averages over the starting point}

In some applications, the precise location of the starting point is
unknown or irrelevant, and it is convenient to average the survival
probability and the PDF of the FPT as if the starting point was
uniformly distributed.  

First, we consider the average over a cross section at a given height
$z$.  We get
\begin{align}  
\overline{\tilde{H}_1}(z) & = \frac{2\pi}{\pi (R_2^2-R_1^2)} \int\limits_{R_1}^{R_2} dr \, r \, \tilde{H}(p|r,z) \\   \nonumber
& = - \frac{2\rho}{1-\rho^2} \biggl(\frac{w'(\rho)}{\alpha^2} +
\sum\limits_{n=0}^\infty c_{n,1} s_{n,1}(z) \frac{v'_{n,1}(\rho)}{\alpha_{n,1}^2}\biggr)  \quad (z < 0),
\end{align}
and
\begin{align}  \label{eq:u2_bar}
\overline{\tilde{H}_2}(z) & = \frac{2\pi}{\pi R_2^2} \int\limits_0^{R_2} dr \, r \, \tilde{H}(p|r,z)  
 = c_{0,2}  \, \frac{\cosh((L_2-z)\sqrt{p/D})}{\cosh(L_2\sqrt{p/D})}   \quad (z > 0),
\end{align}
where we used
\begin{equation}
\int\limits_{\rho}^1 d\r \, \r\, v_{n,1}(\r) = \frac{\rho\, v'_{n,1}(\rho)}{\alpha_{n,1}^2}
\end{equation}
due to the boundary conditions (and similar equation holds for the
integral of $\r w(\r)$), and $\alpha_{0,2}' = \alpha = R_2\sqrt{p/D}$.
One sees that the coefficient $c_{0,2}$ can thus be interpreted as the
cross-sectional average of $\tilde{H}(p|r,z)$ at $z = 0$.  In
addition,
\begin{align} 
\overline{\tilde{H}_1}(-L_1) & = - \frac{2\rho}{1-\rho^2} \biggl(\frac{w'(\rho)}{\alpha^2} 
 + \sum\limits_{n=0}^\infty \frac{c_{n,1} }{\cosh(\alpha_{n,1}' L_1/R_2)} \frac{v'_{n,1}(\rho)}{\alpha_{n,1}^2}\biggr) 
\end{align}
corresponds to the setting when the particle is released from the
bottom surface and has high chances to be absorbed by the pillar; the
knowledge of the PDF allows one to quantify an escape from the
nanoforest of absorbing pillars.  When $L_1/R_2$ is large (i.e., the
pillars are high), the sum can be neglected, and one retrieves the
surface-averaged Laplace-transformed PDF in an annulus between an
absorbing inner circle and a reflecting outer circle.
In contrast, if the particle is released from the top surface,
$\overline{\tilde{H}_2}(L_2)$ characterizes how efficiently the
nanoforest of absorbing pillars can capture such a particle diffusing
from a remote location.

Second, we can use these expressions to compute the volume average,
as if the starting point was uniformly distributed in the bulk:
\begin{align*}
\overline{\tilde{H}} & = \frac{1}{\pi(R_2^2-R_1^2)L_1 + \pi R_2^2 L_2} \biggl(
\pi(R_2^2-R_1^2) \int\limits_{-L_1}^0 dz \overline{\tilde{H}_1}(z) 
 + \pi R_2^2 \int\limits_0^{L_2} dz \overline{\tilde{H}_2}(z)\biggr) \\
& = \frac{1}{(1-\rho^2) h_1 + h_2} \biggl(- 2 \frac{\rho w'(\rho)}{\alpha^2} h_1 
 - 2 \sum\limits_{n=0}^\infty c_{n,1} \frac{\rho v'_{n,1}(\rho)}{\alpha_{n,1}^2} \frac{\tanh(\alpha'_{n,1} h_1)}{\alpha'_{n,1}}  
 + c_{0,2} \frac{\tanh(\alpha h_2)}{\alpha} \biggr),
\end{align*}
where $h_1 = L_1/R_2$ and $h_2 = L_2/R_2$.

\subsection{Long-time behavior in the limit $L_2 = \infty$}
\label{sec:long-time}

In this Section, we discuss the long-time behavior of the PDF
$H(t|r,z)$ in the configuration with $L_2 = \infty$.  We recall that
$L_2$ affects the coefficients $c_{n,1}$ and $c_{n,2}$ of the
Laplace-transformed PDF $\tilde{H}(p|r,z)$ only through the matrix
elements $B_n^{(2)}$ given by Eq. (\ref{eq:Bn2}).  As the long-time
behavior of $H(t|r,z)$ corresponds to the small-$p$ behavior of
$\tilde{H}(p|r,z)$, it is instructive to look at the behavior of
$B_n^{(2)}$ as $p\to 0$.  For $n > 0$, $\alpha_{n,2}' \to \alpha_{n,2}
> 0$, with $O(p)$ corrections, so that the elements $B_n^{(2)}$ tend
to strictly positive limits.  In contrast, $\alpha_{0,2}' =
\alpha = R_2 \sqrt{p/D} \to 0$, and the asymptotic behavior of $B_0^{(2)}$
depends on whether $L_2$ is finite or not:
\begin{equation}
B_0^{(2)} \approx \left\{ \begin{array}{l l} L_2/R_2 & (L_2 < \infty) \\ \alpha & (L_2 = \infty) \\ \end{array} \right.
\quad (p \to 0).
\end{equation}

We start with the case $L_2 = \infty$.  According to the definition
(\ref{eq:Wdef}), the matrix $I+W$ can be written as
\begin{equation}
I + W \approx I + W_0 + \alpha Y + O(p),  
\end{equation}
where $W_0$ denotes the matrix $W$ evaluated at $p = 0$, and the
matrix $Y$ has the elements
\begin{equation}
Y_{k,n} =  2 \delta_{n,0} \sum\limits_{n'=0}^\infty A_{n',k} B_{n'}^{(1)}  A_{n',0}  \,,
\end{equation}
i.e., it has only one nonzero column at $n = 0$.  The coefficients
$c_{n,2}$ can then be found as
\begin{align*}  \nonumber
c_{n,2} & = \bigl[(I+W)^{-1} V]_n 
 \approx \biggl[(I+W_0)^{-1} - \alpha (I+W_0)^{-1} Y (I+W_0)^{-1} + O(p)\biggr]_n .
\end{align*}
Substituting these coefficients into Eq. (\ref{eq:u2}), one gets
\begin{equation}
\tilde{H}(p|r,z) = 1 - C \sqrt{p/D} + O(p),
\end{equation}
where the constant term comes from the normalization, while the
subleading term is of the order of $p^{1/2}$, with some prefactor $C$
(this prefactor can be expressed from the above formulas).  This
asymptotic behavior implies $\tilde{S}(p|r,z) \approx C/\sqrt{pD}$,
from which the Tauberian theorem yields the long-time behaviors:
\begin{equation}  \label{eq:S_longt}
S(t|r,z) \approx \frac{C}{\sqrt{\pi Dt}}   \quad \Rightarrow \quad H(t|r,z) \approx \frac{C}{\sqrt{4\pi Dt^3}} \,.
\end{equation}

The above ``derivation'' does not pretend to be mathematically
rigorous; in fact, one deals here with infinite-dimensional matrices
that requires a more refined analysis, in particular, on the
convergence.  Nevertheless, this derivation highlights the emergence
of the $p^{1/2}$-contribution from the matrix element $B_0^{(2)}$ as
the mathematical origin of the slow power-law decay.  In fact, if
$L_2$ is finite, $B_0^{(2)} \to L_2/R_2$, and there is no
$p^{1/2}$-term.  In this case, one would simply get $\tilde{H}(p|r,z)
= 1 + O(p)$, and the coefficient in front of $-p$ would be the mean
FPT.  Moreover, the analysis of the poles (see \ref{sec:poles}) would
yield the exponential decay of $H(t|r,z)$, in sharp contrast to
Eq. (\ref{eq:S_longt}) for $L_2 = \infty$.

Note that the situation is different in the limit $L_1\to\infty$ (with
a finite $L_2$).  Here, the height $L_1$ affects the coefficients
$c_{n,1}$ and $c_{n,2}$ through the matrix elements $B_n^{(1)}$, which
involve $\alpha_{n,1}'$ that approach strictly positive limits
$\alpha_{n,1}$ as $p\to 0$ for all $n$.  As a consequence, the
$p^{1/2}$-terms do not emerge, and the mean FPT remains finite,
regardless whether $L_1$ is finite or infinite.

\subsection{Poles}
\label{sec:poles}

The poles of the Laplace-transformed survival probability
$\tilde{S}(p|r,z)$ determine the eigenvalues $\lambda_n$ of the
Laplace operator in the considered domain.  As discussed in the text,
the eigenvalues are strictly positive so that all the poles lie on the
negative axis in the complex plane $p\in {\mathbb C}$.  At each pole,
the matrix $I+W$ determining the coefficients $c_{n,2}$ is not
invertible, i.e., its determinant is zero: $\det(I+W) = 0$.  This
equation can be used for a numerical computation of the poles.
However, the computation is rather subtle because the matrix $W$,
which was originally constructed for positive $p$, is divergent at
some negative values of $p$.  We recall that the matrix $W$ depends on
$p$ through two diagonal matrices $B^{(1)}$ and $B^{(2)}$ whose
elements are given by Eqs. (\ref{eq:Bn1}, \ref{eq:Bn2}).  As these
elements involve respectively $\ctanh(\alpha_{n,1}' h_1)$ and
$\tanh(\alpha_{n,2}' h_2)$ (with $h_1 = L_1/R_2$ and $h_2 = L_2/R_2$),
they become infinite when $\alpha_{n,1}' h_1 = i\pi k$ or
$\alpha_{n,2}' h_2 = i(\pi/2 +\pi k)$, for any integer $k$.  In other
words, there are two families of points,
\begin{subequations}
\begin{align}
\frac{R_2^2}{D} p_{n,k}^{(1)} & = - \frac{\pi^2 k^2}{h_1^2} - \alpha_{n,1}^2  \quad 
\left(\begin{array}{l} n=0,1,2,\ldots \\ k = 0,1,2,\ldots \end{array}\right), \\
\frac{R_2^2}{D} p_{n,k}^{(2)} & = - \frac{\pi^2 (k+1/2)^2}{h_2^2} - \alpha_{n,2}^2  \quad 
\left(\begin{array}{l} n=0,1,2,\ldots \\ k = 0,1,2,\ldots \end{array}\right) ,
\end{align}
\end{subequations}
at which $\det(I+W)$ is infinite.  By ordering these points, one can
search for the poles (i.e., the zeros of $\det(I+W)$) on intervals
between each pair of these consecutive points.  These points actually
help to locate the poles.  Moreover, they can also be used to get
upper and lower bounds on each pole.  For instance, the pole $p_0$
determining the principal eigenvalue $\lambda_0$ is bounded by
\begin{equation}
0 < |p_0| \leq \min\bigl\{ |p_{0,0}^{(1)}|, |p_{0,0}^{(2)}|\bigr\} = 
\min\biggl\{ \frac{\alpha_{0,1}^2 D}{R_2^2}, \frac{\pi^2 D}{4L_2^2} \biggr\} ,
\end{equation}
in agreement with the bound (\ref{eq:T_lower}).

\subsection{Thin pillar asymptotic behavior}
\label{sec:Athin}

In this section, we briefly discuss the limit $R_1 \to 0$, which
affects the solutions $\alpha_{n,1}$ of Eq. (\ref{eq:alpha1}) and thus
the matrix elements of $A$ and $B^{(1)}$.  Following a similar
analysis in \cite{Grebenkov23}, we reproduce the asymptotic behavior
(\ref{eq:alpha0_rho0}) of $\alpha_{0,1}$.  In general, $\alpha_{n,1}$
approach $\alpha_{n,2}$ while the associated eigenfunctions
$v_{k,1}(\r)$ approach $\sqrt{2}\, v_{k,2}(\r)$ as $R_1 \to 0$ (see
also \cite{Ward93}).  As a consequence, the matrix $A$, whose elements
were defined in Eq. (\ref{eq:A_def}) as a weighted scalar product of
these functions, approaches $I/\sqrt{2}$, where $I$ is the identity
matrix.  In the leading order, one gets thus
\begin{equation}
W_{k,k'} \approx \delta_{k,k'} B_k^{(1)} B_{k'}^{(2)} \,,
\end{equation}
and the diagonal structure of this matrix allows for the explicit
inversion of $I+W$.  We get therefore
\begin{equation}
c_{n,2} \approx \frac{V_n}{1 + B_n^{(1)} B_n^{(2)}}  \qquad (\rho \to 0),
\end{equation}
where $B_n^{(1)}$ and $B_n^{(2)}$ are given by Eqs. (\ref{eq:Bn1},
\ref{eq:Bn2}).
Using the asymptotic behavior of the modified Bessel functions, we get
in the leading order in $\rho$:
\begin{equation}
w'(\rho) \approx - \frac{\rho^{-1}}{\frac{K_1(\alpha)}{I_1(\alpha)} - \gamma - \ln(\alpha \rho/2)} \,,
\end{equation}
from which
\begin{equation}
V_k \approx \frac{2 v_{k,2}(\rho)}{(\alpha_{k,2}^2 + \alpha^2) \bigl[\frac{K_1(\alpha)}{I_1(\alpha)} - \gamma - \ln(\alpha \rho/2)\bigr]} \,.
\end{equation}
In other words, we obtained a fully explicit approximate solution
which does not require a numerical inversion of the
infinite-dimensional matrix $I+W$.

\section{Auxiliary survival probabilities}

For completeness, we provide here the well-known expressions for the
survival probabilities for one-dimensional and planar diffusions.
When the particle diffuses on the interval $(0,L_2)$ with absorbing
endpoint $0$ and reflecting endpoint $L_2$, the survival probability
reads
\begin{equation} \label{eq:St_interval}
S_{\rm 1D}(t|z) = 2 \sum\limits_{n=0}^\infty \frac{\sin(\pi (n+1/2)z/L_2)}{\pi(n+1/2)} e^{-\pi^2 (n+1/2)^2 Dt/L_2^2}  .
\end{equation} 
In turn, if the particle diffuses in an annulus between an inner
absorbing circle of radius $R_1$ and an outer reflecting circle of
radius $R_2$, the Laplace transform of the PDF is given by
Eq. (\ref{eq:u01}), while its inverse Laplace transform via the
residue theorem yields
\begin{equation} \label{eq:St_annulus}
S_{\rm 2D}(t|r) = \sum\limits_{n=0}^\infty \frac{\rho v'_{n,1}(\rho)}{\alpha_{n,1}^2} v_{n,1}(r/R_2) e^{-\alpha_{n,1}^2 Dt/R_2^2}  .
\end{equation}

\section{Splitting probability}
\label{sec:splitting}

In this Appendix, we consider diffusion in a semi-infinite reflecting
cylindrical tube of radius $R_2$ with a coaxial semi-infinite
absorbing pillar of radius $R_1$: $\Omega = \{ (x,y,z)\in\R^3 ~:~
R_1^2 < x^2 + y^2 < R_2^2, ~ z < 0\}$.  We sketch the computation of
the splitting probability $u(r,z)$, i.e., the probability of hitting
the annular region at the level $z = 0$ before hitting the cylindrical
part of the pillar at $r = R_1$.  The splitting probability satisfies
\begin{subequations}
\begin{align}
\Delta u & = 0  \quad  \textrm{in}~\Omega, \\  
u(R_1,z) & = 0 ,\\    
\label{eq:u_z0}
u(r,0) & = 1, \\
(\partial_r u)(R_2,z) & = 0 , \\
u(r,z) & \to 0 \quad (z \to -\infty).
\end{align}
\end{subequations}
In analogy to the derivation in \ref{sec:exact}, one can search the
solution as
\begin{equation}
u(r,z) = \sum\limits_{n=0}^\infty c_n \, v_{n,1}(r/R_2) \, e^{\alpha_{n,1} z/R_2} \,,
\end{equation}
where the coefficients $c_n$ are found from the boundary condition
(\ref{eq:u_z0}) by multiplication by $\r \, v_{k,1}(\r)$ and
integration over $\r$ from $\rho$ and $1$,
\begin{equation}
c_n = \int\limits_\rho^1 d\r \, \r \, v_{n,1}(\r) = \frac{\rho v'_{n,1}(\rho)}{\alpha_{n,1}^2} .
\end{equation}
When $|z|/R_2$ is large enough, the leading contribution is given by
the first term with the smallest value $\alpha_{0,1}$:
\begin{equation}  \label{eq:u_c0}
u(r,z) \approx C(r) \, e^{\alpha_{0,1} z/R_2} \,,
\end{equation}
with
\begin{equation}  \label{eq:Cr_splitting}
C(r) = \frac{\rho v'_{0,1}(\rho)}{\alpha_{0,1}^2} v_{0,1}(r/R_2) .
\end{equation}


\vskip 10mm
\begin{thebibliography}{100}

\bibitem{Redner}		Redner S 2001
				{\it A Guide to First-Passage Processes}
				(Cambridge University Press)

\bibitem{Rice}			Rice S 1985
				{\it Diffusion-Limited Reactions}
				(Elsevier, Amsterdam)

\bibitem{benAvraham}		ben-Avraham D and Havlin S 2010
				{\it Diffusion and Reactions in Fractals and Disordered Systems}
				(Cambridge University Press)

\bibitem{Metzler}		Metzler R, Oshanin G, and Redner S (Eds) 2014
				{\it First-Passage Phenomena and Their Applications}
				(World Scientific Press, Singapore) 

\bibitem{Lindenberg}		Lindenberg K, Metzler R, and Oshanin G (Eds) 2019
				{\it Chemical Kinetics: Beyond the Textbook}
				(World Scientific, New Jersey)

\bibitem{Condamin07}		Condamin S, B\'enichou O, Tejedor V, Voituriez R, and Klafter J 2007
				First-passage time in complex scale-invariant media
				{\it Nature} {\bf 450} 77

\bibitem{Grebenkov07}		Grebenkov DS 2007 
				NMR Survey of Reflected Brownian Motion 
				{\it Rev. Mod. Phys.} {\bf 79} 1077-1137

\bibitem{Benichou10b}		B\'enichou O, Chevalier C, Klafter J, Meyer B, and Voituriez R 2010
				Geometry-controlled kinetics
				{\it Nature Chem.} {\bf 2} 472-477

\bibitem{Benichou11}		B\'enichou O, Loverdo C, Moreau M, and Voituriez R 2011
				Intermittent search strategies
				{\it Rev. Mod. Phys.} {\bf 83} 81-130

\bibitem{Hofling13}		H\"ofling F and Franosch T 2013
				Anomalous transport in the crowded world of biological cells
				{\it Rep. Progr. Phys.} {\bf 76} 046602

\bibitem{Bressloff13}		Bressloff PC and Newby JM 2013
				Stochastic models of intracellular transport
				{\it Rev. Mod. Phys.} {\bf 85} 135-196

\bibitem{Grebenkov13}		Grebenkov DS and Nguyen B-T 2013 
				Geometrical structure of Laplacian eigenfunctions 
				{\it SIAM Rev.} {\bf 55} 601-667

\bibitem{Grebenkov20}		Grebenkov DS 2020
				Paradigm Shift in Diffusion-Mediated Surface Phenomena
				{\it Phys Rev Lett.} {\bf 125} 078102





\bibitem{Weiss86}		Weiss GH 1986
				Overview of theoretical models for reaction rates
				{\it J. Stat. Phys.} {\bf 42} 3

\bibitem{Benichou08}		B\'enichou O and Voituriez R 2008
				Narrow-Escape Time Problem: Time Needed for a Particle to Exit a Confining Domain through a Small Window
				{\it Phys. Rev. Lett.} {\bf 100} 168105

\bibitem{Holcman14}		Holcman D and Schuss Z 2014
				The Narrow Escape Problem
				{\it SIAM Rev.} {\bf 56} 213-257

\bibitem{Benichou14}		B\'enichou O and Voituriez R 2014
				From first-passage times of random walks in confinement to geometry-controlled kinetics
				{\it Phys. Rep.} {\bf 539} 225-284

\bibitem{Holcman}		Holcman D and Schuss Z 2015 
				{\it Stochastic Narrow Escape in Molecular and Cellular Biology}
				(Springer, New York)

\bibitem{Grebenkov16}		Grebenkov DS 2016
				Universal formula for the mean first passage time in planar domains
				{\it Phys. Rev. Lett.} {\bf 117} 260201

\bibitem{Guerin16}		Gu\'erin T, Levernier N, B\'enichou O, and Voituriez R 2016
				Mean first-passage times of non-Markovian random walkers in confinement
				{\it Nature} {\bf 534} 356-359





\bibitem{Mattos12}		Mattos T, Mej\'{\i}a-Monasterio C, Metzler R, and Oshanin G 2012
				First passages in bounded domains: When is the mean first passage time meaningful? 
				{\it Phys. Rev. E} {\bf 86} 031143

\bibitem{Godec16a}		Godec A and Metzler R 2016 
				Universal proximity effect in target search kinetics in the few encounter limit 
				{\it Phys. Rev. X} {\bf 6} 041037 

\bibitem{Godec16b}		Godec A and Metzler R 2016 
				First passage time distribution in heterogeneity controlled kinetics: going beyond the mean first passage time
				{\it Sci. Rep.} {\bf 6} 20349

\bibitem{Grebenkov18c}		Grebenkov DS, Metzler R and Oshanin G 2018 
				Strong defocusing of molecular reaction times results from an interplay of geometry and reaction control 
				{\it Comm. Chem.} {\bf 1} 96

\bibitem{Reva21}		Reva M, DiGregorio DA, and Grebenkov DS 2021 
				A first-passage approach to diffusion-influenced reversible binding: 
				insights into nanoscale signaling at the presynapse 
				{\it Sci. Rep.} {\bf 11} 5377






\bibitem{Hughes}		Hughes BD 1995 
				{\it Random Walks and Random Environments}
				(Clarendon Press, Oxford) 

\bibitem{Levernier19}		Levernier N, Dolgushev M, B\'enichou O, Voituriez R, and Gu\'erin T 2019
				Survival probability of stochastic processes beyond persistence exponents
				{\it Nature Commun.} {\bf 10} 2990





\bibitem{Kayser83}		Kayser RF and Hubbard JB 1983 
				Diffusion in a Medium with a Random Distribution of Static Traps
				{\it Phys. Rev. Lett.} {\bf 51} 79-82

\bibitem{Kayser84}		Kayser RF and Hubbard JB 1984 
				Reaction diffusion in a medium containing a random distribution of nonoverlapping traps
				{\it J. Chem. Phys.} {\bf 80} 1127-1130

\bibitem{Torquato91}		Torquato S and Avellaneda M 1991 
				Diffusion and reaction in heterogeneous media: pore-size distribution, relaxation times, and mean survival time
				{\it J. Chem. Phys.} {\bf 95} 6477-6489


\bibitem{Levitz06}		Levitz P, Grebenkov DS, Zinsmeister M, Kolwankar KM, and Sapoval B 2006 
				Brownian flights over a fractal nest and first passage statistics on irregular surfaces 
				{\it Phys. Rev. Lett.} {\bf 96} 180601

\bibitem{Lanoiselee18}		Lanoisel\'ee Y, Moutal N, and Grebenkov DS 2018
				Diffusion-limited reactions in dynamic heterogeneous media
				{\it Nature Commun.} {\bf 9} 4398







\bibitem{Basnayake18}		Basnayake K, Hubl A, Schuss Z, and Holcman D 2018
				Extreme narrow escape: Shortest paths for the first particles among n to reach a target window
				{\it Phys. Lett. A} {\bf 382} 3449-3454

\bibitem{Grebenkov22d}		Grebenkov DS, Metzler R and Oshanin G 2022 
				Search efficiency in the Adam-D\"elbruck reduction-of-dimensionality scenario versus direct diffusive search 
				{\it New J. Phys.} {\bf 24} 083035


\bibitem{Varadhan67a}		Varadhan SRS 1967
				On the Behavior of the Fundamental Solution of the Heat Equation with Variable Coefficients
				{\it Comm. Pure. Appl. Math.} {\bf 20} 431-455

\bibitem{Varadhan67b}		Varadhan SRS 1967
				Diffusion Processes in a Small Time Interval
				{\it Comm. Pure. Appl. Math.} {\bf 20} 659-685

\bibitem{Smith19}		Smith NR and Meerson B 2019
				Geometrical optics of constrained Brownian excursion: from the KPZ scaling to dynamical phase transitions
				{\it J. Stat. Mech.} 023205

\bibitem{Meerson22}		Meerson B and Oshanin G 2022
				Geometrical optics of large deviations of fractional Brownian motion
				{\it Phys. Rev. E} {\bf 105} 064137




\bibitem{Carslaw}		Carslaw HS and Jaeger JC 1959
				{\it Conduction of Heat in Solids}, 2nd Ed.
				(Oxford University Press)

\bibitem{Crank}			Crank J 1956
				{\it The Mathematics of Diffusion}
				(Oxford University Press)

\bibitem{Thambynayagam}		Thambynayagam RKM 2011
				{\it The Diffusion Handbook: Applied Solutions for Engineers}
				(New York: McGraw-Hill Education)




\bibitem{Isaacson13}		Isaacson SA and Newby J 2013
				Uniform asymptotic approximation of diffusion to a small target
				{\it Phys. Rev. E} {\bf 88} 012820




\bibitem{Rupprecht15} 		Rupprecht J-F, B\'enichou O, Grebenkov DS, and Voituriez R 2015 
				Exit time distribution in spherically symmetric two-dimensional domains 
				{\it J. Stat. Phys.} {\bf 158} 192-230






\bibitem{Grebenkov18d}		Grebenkov DS, Metzler R and Oshanin G 2018 
				Towards a full quantitative description of single-molecule reaction kinetics in biological cells 
				{\it Phys. Chem. Chem. Phys.} {\bf 20} 16393

\bibitem{Grebenkov19c}		Grebenkov DS, Metzler R and Oshanin G 2019 
				Full distribution of first exit times in the narrow escape problem 
				{\it New J. Phys.} {\bf 21} 122001

\bibitem{Grebenkov21c}		Grebenkov DS, Metzler R, and Oshanin G 2021 
				Distribution of first-reaction times with target sites on boundaries of shell-like regions
				{\it New J. Phys.} {\bf 23} 123049

\bibitem{Grebenkov20d}		Grebenkov DS 2020 
				Diffusion toward non-overlapping partially reactive spherical traps: fresh insights onto classic problems 
				{\it J. Chem. Phys.} {\bf 152} 244108





\bibitem{Grebenkov23}		Grebenkov DS and Skvortsov AT 2022
				Diffusion towards a nanoforest of absorbing pillars 
				{\it J. Chem. Phys.} {\bf 157} 244102




\bibitem{Grebenkov18}		Grebenkov DS and Krapf D 2018
				Steady-state reaction rate of diffusion-controlled reactions in	sheets 
				{\it J. Chem. Phys.} {\bf 149} 064117

\bibitem{Delitsyn18}		Delitsyn A and Grebenkov DS 2018 
				Mode matching methods in spectral and scattering problems 
				{\it Quart. J. Mech. Appl. Math.} {\bf 71} 537-580

\bibitem{Delitsyn22}		Delitsyn A and Grebenkov DS 2022 
				Resonance scattering in a waveguide with identical thick perforated barriers 
				{\it Appl. Math. Comput.} {\bf 412} 126592



\bibitem{Keller_1967}           Keller KH and Stein TR 1967
				A Two-Dimensional Analysis of Porous Membrane Transport
				{\it Math. Biosci.} {\bf 1} 421-437




\bibitem{Cai92}			Cai X and Wallis GB 1992
				Potential flow around a row of spheres in a circular tube
				{\it Phys. Fluids A} {\bf 4} 904




\bibitem{Yuste13}		Yuste SB, Abad E, and Lindenberg K 2013
				Exploration and Trapping of Mortal Random Walkers
				{\it Phys. Rev. Lett.} {\bf 110} 220603

\bibitem{Meerson15}		Meerson B and Redner S 2015
				Mortality, Redundancy, and Diversity in Stochastic Search
				{\it Phys. Rev. Lett.} {\bf 114} 198101

\bibitem{Grebenkov17d}		Grebenkov DS and Rupprecht J-F 2017 
				The escape problem for mortal walkers 
				{\it J. Chem. Phys.} {\bf 146} 084106

\bibitem{Meerson19}		Meerson B 2019 
				Mortal Brownian motion: Three short stories
				{\it Int. J. Modern Phys. B} {\bf 33} 1950172



\bibitem{Talbot79}		Talbot A 1979
				The Accurate Numerical Inversion of Laplace Transforms 
				{\it IMA J. Appl. Math.} {\bf 23} 97-120




\bibitem{Mazya85}		Maz'ya VG, Nazarov SA, and Plamenevskii BA 1985
				Asymptotic Expansions of the Eigenvalues of Boundary Value Problems for 
				the Laplace Operator in Domains with Small Holes
				{\it Math. USSR. Izv.} {\bf 24} 321-345

\bibitem{Ward93}		Ward MJ and Keller JB 1993
				Strong Localized Perturbations of Eigenvalue Problems
				{\it SIAM J. Appl. Math.} {\bf 53} 770-798

\bibitem{Kolokolnikov05}	Kolokolnikov T, Titcombe MS, and Ward MJ 2005
				Optimizing the Fundamental Neumann Eigenvalue for the Laplacian in a Domain with Small Traps
				{\it Eur. J. Appl. Math.} {\bf 16} 161

\bibitem{Cheviakov11}		Cheviakov AF and Ward MJ 2011
				Optimizing the principal eigenvalue of the Laplacian in a sphere with interior traps
				{\it Math. Computer Model.} {\bf 53} 1394-1409

\bibitem{Chaigneau22}		Chaigneau A and Grebenkov DS 2022 
				First-passage times to anisotropic partially reactive targets 
				{\it Phys. Rev. E} {\bf 105} 054146



\bibitem{Sandua_2013}           Sandua T, Boldeiu G, and Moagar-Poladian V 2013
				Applications of electrostatic capacitance and charging
				{\it J. Appl. Phys.} {\bf 114} 224904






\bibitem{Berg77}		Berg HC and Purcell EM 1977
				Physics of chemoreception
				{\it Biophys. J.} {\bf 20} 193

\bibitem{Berezhkovskii07}	Berezhkovskii AM and Barzykin AV 2007
				Simple formulas for the trapping rate by nonspherical absorber and capacitance of nonspherical conductor
				{\it J. Chem. Phys.} {\bf 126} 106102

\bibitem{Lindsay_2017}		Lindsay AE, Bernoff AJ, and Ward MJ 2017
				First passage statistics for the capture of a Brownian particle by a 
				structured spherical target with multiple surface traps
				{\it SIAM Multiscale Model. Simul.} {\bf 15}, 74-109

\bibitem{Grebenkov22c}		Grebenkov DS and Skvortsov AT 2022 
				Mean first-passage time to a small absorbing target in three-dimensional elongated domains 
				{\it Phys. Rev. E} {\bf 105} 054107




\bibitem{Berezhkovskii_2004}	Berezhkovskii AM, Makhnovskii YA, Monine MI, Zitserman VYu, and Shvartsman SY 2004
				Boundary homogenization for trapping by patchy surfaces
				{\it J. Chem. Phys.} {\bf 121} 11390

\bibitem{Bernoff18}		Bernoff AJ, Lindsay AE, and Schmidt DD 2018
				Boundary Homogenization and Capture Time Distributions of Semipermeable Membranes with 
				Periodic Patterns of Reactive Sites
				{\it SIAM Multiscale Model. Simul.} {\bf 16} 1411-1447

\bibitem{Bernoff18b}		Bernoff AJ and Lindsay AE 2018
				Numerical approximation of diffusive capture rates by planar and spherical surfaces with absorbing pores
				{\it SIAM J. Appl. Math.} {\bf 78} 266-290





\bibitem{Morters}       	M\"orters P and Peres Y 2010
                        	{\it Brownian Motion}
                        	(Cambridge Series in Statistical and Probabilistic Mathematics, Cambridge University Press, New York)



\bibitem{Koplik94}		Koplik J, Redner S, and Hinch EJ 1994
				Tracer dispersion in planar multipole flows
				{\it Phys. Rev. E} {\bf 50} 4650

\bibitem{Koplik95}		Koplik J, Redner S, and Hinch EJ 1995
				Universal and Nonuniversal First-Passage Properties of Planar Multipole Flows
				{\it Phys. Rev. Lett.} {\bf 74} 82

\bibitem{Levitz08}		Levitz PE, Zinsmeister M, Davidson P, Constantin D, and Poncelet O 2008
				Intermittent Brownian dynamics over a rigid strand: Heavily tailed relocation statistics
				{\it Phys. Rev. E} {\bf 78} 030102(R)

\bibitem{Grebenkov21}		Grebenkov DS 2021 
				Statistics of boundary encounters by a particle diffusing outside a compact planar domain 
				{\it J. Phys. A.: Math. Theor.} {\bf  54} 015003







\bibitem{Kharisov_2015}         Kharisov BI, Kharissova OV, Garc{\'\i}a BO, M\'endez YP, and de la Fuente IG 2015
				State of the art of nanoforest structures and their applications
				{\it Proc. Roy. Soc. Adv.} {\bf 5}, 105507

\bibitem{Davis_2010}            Davis AMJ and Lauga E 2010
				Hydrodynamic friction of fakir--like superhydrophobic surfaces
				{\it J. Fluid Mech.} {\bf 661} 402-411

\bibitem{Ramon_2012}        	Ramon GZ, Wong MCY, Hoek EMV 2012
				Transport through composite membrane, part 1: Is there an optimal support membrane? 
				{\it J. Membr. Sci.} {\bf 415-416} 298-305
        
\bibitem{Ramon_2013}		Ramon GZ and Hoek EMV 2013
				Transport through composite membranes, part 2: Impacts of roughness on permeability and fouling
				{\it J. Membr. Sci.} {\bf 425-426} 141-148


\bibitem{Nair_2007}             Nair PR and Alam MA 2007
				Dimensionally frustrated diffusion towards fractal adsorber
				{\it Phys. Rev. Lett.} {\bf 99} 256101

                            
\bibitem{Chen_2022}             Chen G, Guan R, Shi M, Dai X, Li H, Zhou N, Chen D, and Mao H 2022
				A nanoforest-based humidity sensor for respiration monitoring
				{\it Microsystems and Nanoeng.} {\bf 8} 44
   
\bibitem{Borberg19}		Borberg E et al. 2019 
				Light-Controlled Selective Collection-and-Release of Biomolecules by an On-Chip Nanostructured Device
				{\it NanoLett.} {\bf 19} 5868-5878





\bibitem{Watson}		Watson GN 1962 
				{\it A Treatise on the Theory of Bessel Functions}
				(Cambridge University Press, Cambridge)



\end{thebibliography}
\end{document}